\documentclass[reprint,amsmath,amssymb,aip]{revtex4-2}

\usepackage{graphicx}
\usepackage{dcolumn}
\usepackage{bm}

\begin{document}

\title{Electric control of optically-induced magnetization dynamics in a van der Waals ferromagnetic semiconductor}

\author{Freddie Hendriks}
\affiliation{Zernike Institute for Advanced Materials, University of Groningen, The Netherlands}

\author{Rafael R. Rojas-Lopez}
\affiliation{Zernike Institute for Advanced Materials, University of Groningen, The Netherlands}
\affiliation{Departamento de F\'{i}sica, Universidade Federal de Minas Gerais, Brazil}

\author{Bert Koopmans}
\affiliation{Department of Applied Physics, Eindhoven University of Technology, The Netherlands}

\author{Marcos H. D. Guimar\~{a}es}
\affiliation{Zernike Institute for Advanced Materials, University of Groningen, The Netherlands}
\email{m.h.guimaraes@rug.nl}

\date{22 September 2023}
\begin{abstract}
Electric control of magnetization dynamics in two-dimensional (2D) magnetic materials is an essential step for the development of novel spintronic nanodevices.
Electrostatic gating has been shown to greatly affect the static magnetic properties of some van der Waals magnets, but the control over their magnetization dynamics is still largely unexplored.
Here we show that the optically-induced magnetization dynamics in the van der Waals ferromagnet Cr$_2$Ge$_2$Te$_6$ can be effectively controlled by electrostatic gates, with a one order of magnitude change in the precession amplitude and over 10\% change in the internal effective field. 
In contrast to the purely thermally-induced mechanisms previously reported for 2D magnets, we find that coherent opto-magnetic phenomena play a major role in the excitation of magnetization dynamics in Cr$_2$Ge$_2$Te$_6$.
Our work sets the first steps towards electric control over the magnetization dynamics in 2D ferromagnetic semiconductors, demonstrating their potential for applications in ultrafast opto-magnonic devices.
\end{abstract}


\maketitle

\section{Introduction}
Ever since the experimental confirmation of magnetism in two-dimensional (2D) van der Waals (vdW) materials \cite{Huang2017, Gong2017}, researchers have tried to understand their fundamentals and to utilise their unique properties for new technologies, such as novel spintronic devices for information storage and processing \cite{Gong2019, Ningrum2020, Huang2020, Wang2022}.
The use of magnetization dynamics is particularly interesting since it provides an energy efficient route to transfer and process information \cite{Cornelissen2015, Chumak2015, Lebrun2018, Manipatruni2019, Puebla2020}.
A key challenge in this field, named magnonics, is the effective control over the magnetization and its dynamics using electrostatic means, allowing for energy efficient, on-chip, reconfigurable magnonic circuit elements \cite{Wang2017, Chumak2017, Merbouche2021}.
For conventional (tree-dimensional) systems this control has been shown to be very promising to reduce the energy barriers for writing magnetic bits using spin-orbit torques \cite{Fan2016, Jiang2023}.
Nonetheless, the effect is still relatively modest \cite{Weisheit2007, endo2010, chiba2011, Dietl2014}.
In contrast, 2D magnetic semiconductors provide an ideal platform for electric manipulation of magnetization.
Their low density of states and high surface-to-volume ratio allow for an effective control over the magnetic parameters in these systems, such as the magnetic anisotropy and saturation magnetization \cite{Wang2018a, Huang2018, Jiang2018, Verzhbitskiy2020, Zhuo2021}.
Additionally, 2D magnetic semiconductors offer a bridge to another exciting field: the combination of optics and magnetism.
These materials have shown to possess strong light-matter interaction and high magneto-optic coefficients which strength can be further tuned by the use of vdW heterostructures \cite{Fang2018, Wu2019, KumarGudelli2019, Molina-Sanchez2020, Yang2021, Wang2022, Hendriks2021}.
These properties make 2D magnetic semiconductors ideal for the merger of two emerging fields: magnonics and photonics.

Most works on the electric control of magnetization in vdW magnets have focused on their magnetostatic properties, such as the magnetic anisotropy, saturation magnetization and Curie temperature, in both metallic \cite{Deng2018,Zheng2020, Chen2021, Verzhbitskiy2020, Zhuo2021, Wu2023} and semiconducting \cite{Huang2018, Jiang2018, Wang2018a, Verzhbitskiy2020, Zhuo2021} materials.
In contrast, their magnetization dynamics have only recently started to receive more attention, and studied using microwave driven magnetic resonance \cite{Ni2021, Zeisner2020, Macneill2019, Zeisner2019, Khan2019, Zollitsch2022, Xu2023}, or time-dependent magneto-optic techniques \cite{Gong2022, Lichtenberg2023, Zhang2020, Sun2021, Sutcliffe2023, Bae2022, Cho2023, Afanasiev2021, Khusyainov2023}.
The latter were used on antiferromagnetic bilayer CrI$_\text{3}$ to show that its magnetic resonance frequency can be electrically tuned by tens of GHz \cite{Zhang2020a}.
Nonetheless, the electric control over the optical excitation of magnetization and its subsequent dynamics in 2D ferromagnets remains to be explored.

Here we show that the magnetization dynamics of the vdW semiconductor Cr$_{2}$Ge$_{2}$Te$_{6}$ (CGT) can be efficiently controlled by electrostatic gating.
Using ultrafast (fs) laser pulses we bring the magnetization out of equilibrium and study its dynamics with high temporal resolution through the magneto-optic Faraday effect.
Using both top and bottom electrostatic gates, we independently control the gate-induced change in the charge carrier density ($\Delta n$) and the electric displacement  field ($\Delta D$) in the CGT, and show that both have drastic effects on the optically-induced oscillation amplitudes and a more modest effect on its frequency.
Finally, we observe a strong asymmetric behavior on the magnetization oscillation amplitudes with respect to a reversal of the external magnetic field, which is also strongly affected by both $\Delta n$ and $\Delta D$.
This asymmetry can be explained by a strong influence of coherent opto-magnetic phenomena, such as the inverse Cotton-Mouton effect and photo-induced magnetic anisotropy, on the excitation of the magnetization dynamics.

\begin{figure*}
	\centering
	\includegraphics[width=\textwidth]{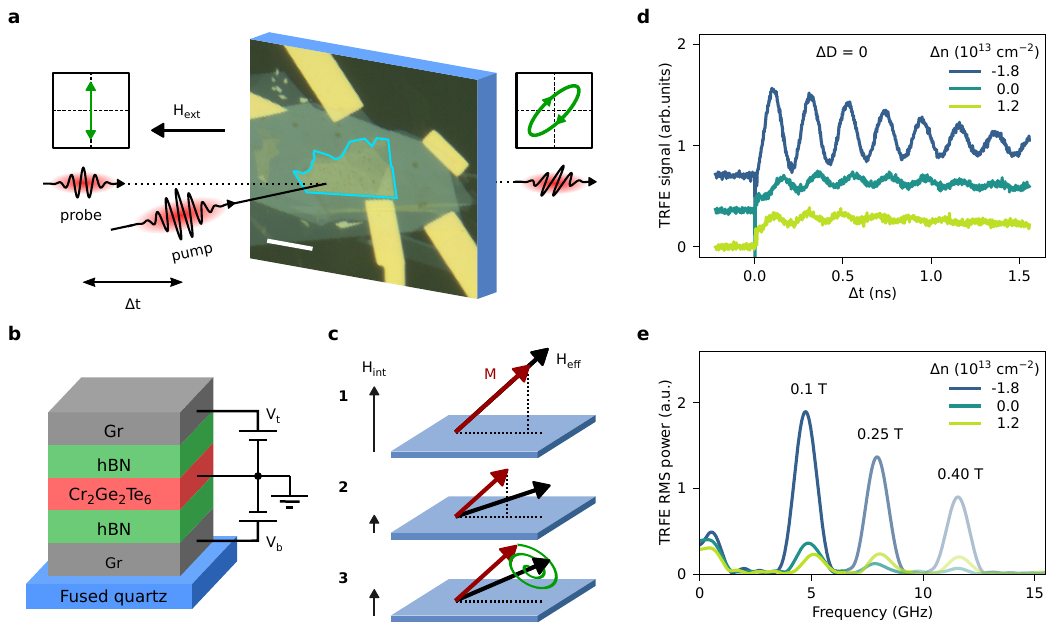}
	\caption{\textbf{Magnetization dynamics in a CGT based heterostructure.}
    \textbf{a}, Illustration of time-resolved Faraday ellipticity measurements, combined with an optical micrograph of the sample (the scale bar is 10 $\mu$m). The CGT flake is outlined in blue. 
    \textbf{b}, Schematic of the layers comprising the sample, including electrical connections for gating.
    \textbf{c}, Process of laser-induced magnetization precession (see main text).
    \textbf{d}, Time-resolved Faraday ellipticity traces at $\mu_0 H_\text{ext}$ = 100 mT for three different values of $\Delta n$ with $\Delta D$ = 0. A vertical offset was added for clarity.
    \textbf{e}, RMS power of the frequency spectrum of the oscillations in the data shown in \textbf{d}. Different transparencies indicate different values of $H_\text{ext}$.
    }
	\label{fig:1}
\end{figure*}

\section{Device structure and measurement techniques}
Our sample consists of a CGT flake, encapsulated in hexagonal boron nitride (hBN), with thin graphite layers as top gate, back gate, and contact electrodes, as depicted in Fig. \ref{fig:1}a and b (see Methods for more details).
The measurements were performed at low temperatures (10 K), with the sample mounted at 50 degrees with respect to the magnetic field axis for transmission measurements.
The laser light is parallel to the magnetic field axis.

We use the time-resolved magneto-optic Faraday effect to monitor the magnetization dynamics in our system using a single-color pump-probe setup similar to the one described in \cite{Guimaraes2018, Rojas-Lopez2023} (more information in Methods).
The process of optical excitation of magnetization dynamics in van der Waals magnets has been previously reported as purely thermal \cite{Gong2022, Lichtenberg2023, Zhang2020, Sun2021, Sutcliffe2023, Zhang2020a}, similar to many studies on conventional metallic thin-films \cite{Koopmans2010, Kim2022, Scheid2022}. 
Here we find strong evidence that coherent opto-magnetic phenomena also play an important role in the excitation of the magnetization dynamics.
The detailed microscopic description of how the magnetization dynamics is induced is described later in the article, but in short, the excitation of the magnetization dynamics can be described as follows (Fig. \ref{fig:1}c):
In equilibrium (1), the magnetization $\mathbf{M}$ points along the total effective magnetic field $\mathbf{H}_\text{eff}$, which is the sum of the external field ($\mathbf{H}_\text{ext}$), and the internal effective field ($\mathbf{H}_\text{int}$) caused by the magnetocrystalline anisotropy ($K_\text{u}$) and shape anisotropy.
For CGT, $\mathbf{H}_\text{int}$ points out-of-plane \cite{Carteaux1995,Gong2017, Zhang2020}, meaning that $K_\text{u}$ dominates over the shape anisotropy.
The linearly polarized pump pulse interacts with the sample (2), reducing the magnetization and changing the magnetocrystalline anisotropy through the mechanisms mentioned above, which causes $\mathbf{M}$ to cant away from equilibrium.
Since $\mathbf{M}$ and $\mathbf{H}_\text{eff}$ are not parallel anymore, this results in a precession of $\mathbf{M}$ around $\mathbf{H}_\text{eff}$, while they both recover to their equilibrium value as the sample cools.

\section{Gate control of magnetization dynamics}
The dual-gate geometry of our device allows for the independent control of both the charge carrier density and the perpendicular electric field.
The dependence of $\Delta n$ and $\Delta D$ on the top and back gate voltages -- $V_t$ and $V_b$, respectively -- is derived in the Methods.
The change in the Fermi level induced by $\Delta n$ is expected to affect the magnetic anisotropy of CGT due to the different Cr $d$-orbitals composition of the electronic bands \cite{Ren2021}.
The effect of $\Delta D$ is, however, more subtle.
The inversion symmetry breaking caused by $\Delta D$ can allow for an energy shift of the (initially degenerate) electronic bands, potentially also modulating the magnetization parameters.
Additionally, the perpendicular electric field can induce a non-uniform distribution of charge carriers along the thickness of the CGT flake, leading to $\Delta n$-induced local changes in the magnetization parameters.

Typical results from the time-resolved Faraday ellipticity (TRFE) measurements for different values of $\Delta n$, with $\Delta D$ = 0, are shown in Fig. \ref{fig:1}d.
For $\Delta t < 0$ the signal is constant, since the magnetization is at its steady state value. 
All traces show a sharp increase at $\Delta t = 0$, indicating a fast laser-induced dynamics.
For $\Delta t > 0$, the TRFE traces show oscillations, indicating a precession of the magnetization induced by the pump pulse.

We observe that the magnetization dynamics strongly depends on $\Delta n$, with the amplitude, frequency and starting phase of the oscillations in the TRFE signal all being affected.
The most striking observation is that the amplitude of the TRFE signal increases by more than a factor of seven when $\Delta n$ is changed from $1.2\times10^{13}$ cm$^{-2}$ to $-1.8\times 10^{13}$ cm$^{-2}$.
The observations of modulation of both the amplitude and starting phase of the oscillations hint at a change in the pump excitation process.
The change in oscillation frequency due to $\Delta n$ is better visible in the Fourier transform of the signals, shown in Fig. \ref{fig:1}e (see Methods for details on the Fourier transform).
This analysis clearly shows that both the frequency and amplitude of the magnetization precession are tuned by $\Delta n$.
All these observations point to an effective control of the (dynamic) magnetic properties of CGT by electrostatic gating.

\begin{figure*}
	\centering
	\includegraphics[width=\textwidth]{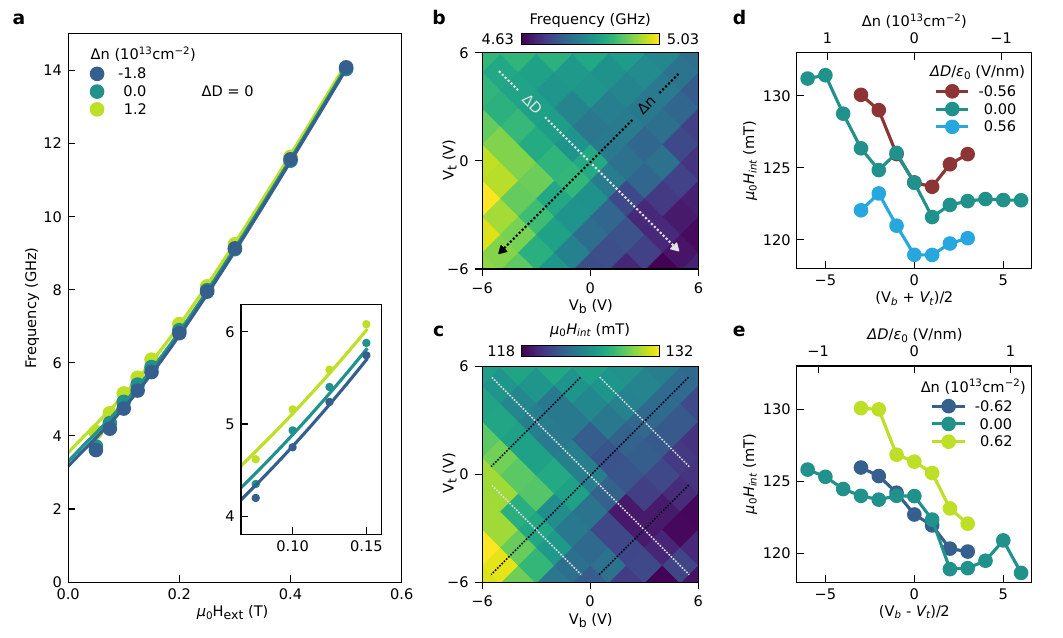}
	\caption{\textbf{Gate-dependence of precession frequency and internal effective field.}
    \textbf{a}, Frequency of oscillations as a function of external magnetic field, for different values of $\Delta n$. The circles are the frequencies extracted from the TRFE data for $\Delta t > 26$ ps. Solid lines are best fits of Eq (\ref{eq:frequency_simple}). \textit{Inset}: Close-up of the data for low fields, showing the frequency shift due to gating. The error bars are smaller than the markers.
    \textbf{b}, Frequency of the oscillations in the TRFE signal at $\mu_0 H_\text{ext} =  100$ mT for various values of top ($V_t$) and back gate voltages ($V_b$). The black and gray arrows indicate, respectively, the directions of constant $\Delta D$ and varying $\Delta n$, and of constant $\Delta n$ and varying $\Delta D$. For other values of $H_\text{ext}$ see Supplementary Fig. S10.
    \textbf{c}, Internal effective field as a function of $V_t$ and $V_b$.
    \textbf{d}, \textbf{e}, The dependence of the internal effective field on $\Delta n$ for fixed $\Delta D$ (\textbf{d}) and on $\Delta D$ for fixed $\Delta n$ (\textbf{e}), with solid lines to guide the eye. The traces are taken along the dotted lines indicated in \textbf{c}.
   }
	\label{fig:2}
\end{figure*}

The origin of the electric control of the magnetization dynamics can be further understood by analyzing the precession frequency at various magnetic fields and values of $\Delta n$ (Fig. \ref{fig:2}a).
For magnetic fields below 250 mT we observe a significant shift of the frequency (4 -- 10\%) by changing the charge carrier density.
This is clearly visible in the inset of Fig. \ref{fig:2}a, which shows a close-up of the data up to 150 mT.
The change in precession frequency for different values of $\Delta n$ strongly points towards a modulation of the magnetization parameters of CGT as a function of the Fermi level, controlled by $\Delta n$.

A quantitative analysis of the oscillation frequency ($f$) as a function of $H_\text{ext}$ can be used to extract the magnetization dynamics parameters of the device.
Our data is well described by the ferromagnetic resonance mode obtained from the Landau-Lifshitz-Gilbert (LLG) equation with negligible damping \cite{Mizukami2010}:
\begin{align}
	f = \frac{g \mu_B \mu_0}{2\pi\hbar} \sqrt{\left|\mathbf{H}_\text{eff}\right|\left(\left|\mathbf{H}_\text{eff}\right| - H_\text{int} \sin^2\left(\theta_M\right)\right)},
    \label{eq:frequency_simple}
\end{align}

\noindent where $g$ is the Land\'{e} g-factor, $\mu_B$ the Bohr magneton, $\mathbf{H}_\text{eff} = \mathbf{H}_\text{ext} + H_\text{int}\cos(\theta_M)\hat{\mathbf{z}}$, with $H_\text{int} = 2K_\text{u}/(\mu_0 M_\text{s}) - M_\text{s}$, $M_\text{s}$ the saturation magnetization, and $\theta_M$ the angle between $\mathbf{M}$ and the sample normal (z-direction).
The angle $\theta_M$ is calculated by minimizing the magnetic energy in the presence of an external field, perpendicular magnetic anisotropy, and shape anisotropy \cite{Zhang2020}.
We obtain the $g$-factor and $H_\text{ext}$ of the CGT by fitting the $f$ versus $H_\text{ext}$ data (e.g. the data presented in Fig. \ref{fig:2}a) using Eq. (\ref{eq:frequency_simple}), as explained in the Methods.
This yields $g \approx 1.89$ with no clear dependence on $\Delta n$ or $\Delta D$, which is in agreement with (albeit slightly lower than) the values reported for CGT \cite{Khan2019, Zhang2020, Xu2023} (see Supplementary Sections 4 and 5 for more details).
We also find no clear dependence of the precession damping time ($\tau_\text{osc}$) on $\Delta n$ or $\Delta D$.
The intrinsic Gilbert damping we obtain form our measurements is about $6\times10^{-3}$ (see Supplementary Section 6), in line with values found in literature \cite{Zhang2020, Zollitsch2022}.

The internal effective field shows a clear dependence on both $V_t$ and $V_b$, as shown in Fig. \ref{fig:2}c, with values similar to the ones found in other studies \cite{Zhang2020}.
Upon comparing Fig. \ref{fig:2}c to \ref{fig:2}b, one notices that the gate dependence of $H_\text{int}$ is very similar to that of the precession frequency at $\mu_0 H_\text{ext} = 100$ mT.
This suggests that the gate dependence of the precession frequency is caused by the gate-induced change in $H_\text{int}$.
From the dependence of $H_\text{int}$ on $V_t$ and $V_b$, we extract its behavior as a function of $\Delta n$ and $\Delta D$, shown in Fig. \ref{fig:2}d and e.
We observe that $H_\text{int}$ decreases with both increasing $\Delta n$ and $\Delta D$.
The dependence of $H_\text{int}$ on $\Delta n$ is consistent with theoretical calculations \cite{Ren2021}, 
showing that $K_\text{u}$, and therefore $H_\text{int}$, is reduced upon increasing the electron density in the same order of what we achieve in our sample.
The $\Delta n$ dependence of $H_\text{int}$ is also consistent with the dependence of the coercive field obtained from static measurements (see Supplementary Section 3), providing further evidence that the change in $f$ is driven by a change in $K_\text{u}$.

\begin{figure*}
	\centering
	\includegraphics[width=\textwidth]{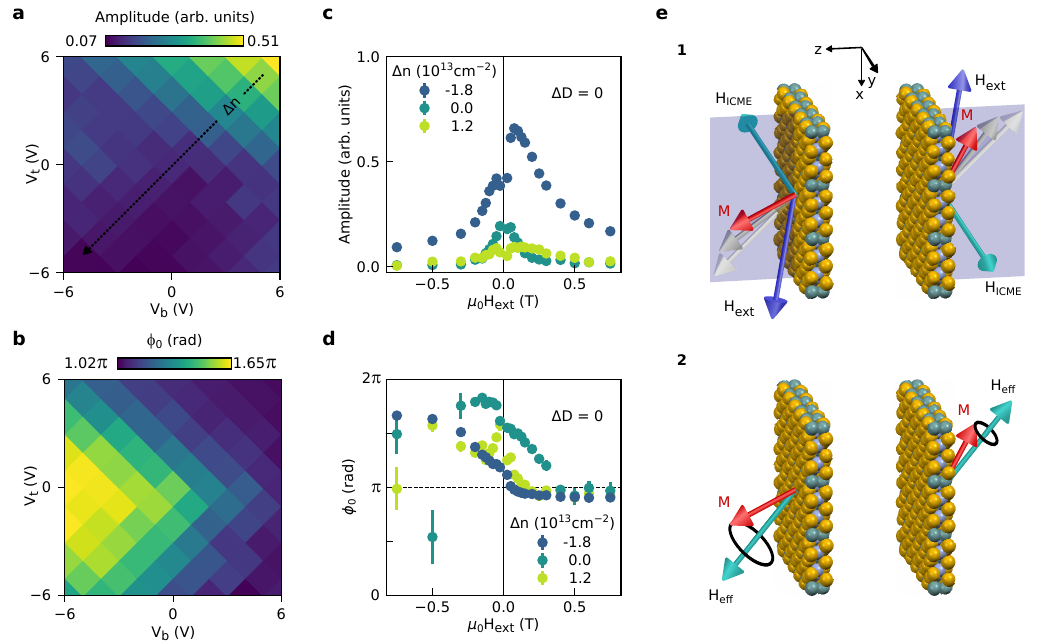}
	\caption{\textbf{Gate dependence of magnetization precession amplitude and phase.} \textbf{a}, \textbf{b}, Gate dependence of the amplitude (\textbf{a}) and starting phase (\textbf{b}) of the oscillations in the TRFE measurements at $\mu_0 H_\text{ext} = 100$ mT. For other values of $H_\text{ext}$ see Supplementary Figs. S8 and S9. \textbf{c}, \textbf{d}, External magnetic field dependence of the amplitude (\textbf{c}) and starting phase (\textbf{d}) of the oscillations for different values of $\Delta n$ at $\Delta D = 0$. The values are extracted from the TRFE data for $\Delta t > 26$ ps.
    \textbf{e}, Schematics of the inverse Cotton-Mouton effect for opposite directions of $H_\text{ext}$.
    The magnetization direction is depicted by a red arrow, the external magnetic field in blue, the effective magnetic field induced by the ICME and the effective field are shown in cyan. The $xz$-plane is highlighted by the shaded region.
    }
    \label{fig:3}
\end{figure*}

Now we draw our attention to the large modulation of the oscillations in the TRFE measurements with varying gate voltage, as shown in Fig. \ref{fig:1}d and \ref{fig:1}e.
Here we attribute this change in magneto-optical signal amplitude to an actual increase in amplitude of the magnetization precession (increase in the precession angle) and not to an increase in the strength of the Faraday effect.
This is supported by our observation that the time-resolved measurements for different combinations of gate voltages are not simply scaled -- i.e. the amplitude of the oscillations and their (ultrafast demagnetization) background scale differently.
A detailed discussion can be found in Supplementary Section 7.

Figure \ref{fig:3}a clearly shows that the magnetization precession amplitude is mostly affected by $\Delta n$, and to a much lesser extend by $\Delta D$.
The precession amplitude versus $H_\text{ext}$ for various values of $\Delta n$ is presented in Fig. \ref{fig:3}c.
We note that for $|\mu_0 H_\text{ext}| < $ 50 mT the magnetization is not completely saturated (see Supplementary Section 3), which can lead to multi-domain formation \cite{Lohmann2019} and a deviation from the general trend.
We find that not only the precession amplitude for a given $H_\text{ext}$ is strongly modulated by $\Delta n$, but its decaying trend with $H_\text{ext}$ is also strongly affected.
Additionally, we observe another interesting effect: the amplitude shows an asymmetry in the sign of the applied magnetic field, which is also dependent on $\Delta n$.
This latter is unexpected, especially since the observed precession frequency is symmetric in H$_\text{ext}$ (see Supplementary Section 9).
A similar precession frequency for opposite magnetic fields indicates that the magnetocrystalline anisotropy and the saturation magnetization are independent of the sign of $H_\text{ext}$.
Therefore, we conclude that the origin of the modulation of the precession amplitude is related to the excitation mechanism of the magnetization precession (see Supplementary Section 10 for the complete discussion).

To get further insight into the microscopic mechanisms involved in the optical excitation of magnetization dynamics, we analyze the magnetic field dependence of the starting phase ($\phi_0$) of the precessions for different values of $V_t$ and $V_b$ (Fig. \ref{fig:3}b).
Unlike the amplitude, we find that $\phi_0$ depends on both $\Delta n$ and $\Delta D$.
As can be seen in Fig. \ref{fig:3}d, the behavior of $\phi_0$ with $H_\text{ext}$ is also modulated by $\Delta n$.
For a purely thermal excitation of the magnetization dynamics one would expect $\phi_0 (-H_\text{ext}) = \pi + \phi_0 (H_\text{ext})$ in our geometry.
Nonetheless, we observe that $\phi_0$ for positive and negative magnetic fields differ by less than $\pi$.
Moreover, $\Delta n$ seems to also affect the trend on how $\phi_0$ approaches the values at high magnetic fields.
Combined with the observed asymmetry of the precession amplitude, our data strongly suggests that the optical excitation of the magnetization dynamics is not dominated by a thermal excitation ($\Delta$K mechanism) as previously reported for other van der Waals magnets  \cite{Gong2022, Lichtenberg2023, Zhang2020, Sun2021, Sutcliffe2023, Zhang2020a}.

\section{Opto-magnetic effects}
Coherent opto-magnetic mechanisms provide possible alternatives for the optical excitation of magnetization dynamics in CGT.
Here we find that our data can be explained by two of these mechanisms that are compatible with a linearly-polarized pump pulse, the inverse Cotton-Mouton effect (ICME) \cite{Pershan1966, Gridnev2008, Kalashnikova2008} and photo-induced magnetic anisotropy (PIMA) \cite{Hansteen2005}, in addition to the conventional (thermal) $\Delta$K mechanism \cite{VanKampen2002}.
The ICME, which could be described by impulsive stimulated Raman scattering, relies on the generation of an effective magnetic field upon interaction with linearly polarized light in a magnetized medium \cite{Pershan1966, Yan1985,Gridnev2008, Kalashnikova2008, Kirilyuk2010}.
This effective magnetic field is proportional to both the light intensity and magnetization.
For pulsed laser excitation, the ICME generates a strong impulsive change in $H_\text{ext}$ that results in a fast rotation of the magnetization.
Therefore, this effect can cause the amplitude of the precession to be asymmetric in $H_\text{ext}$ \cite{Kalashnikova2008, Kirilyuk2010}.

Figure \ref{fig:3}e illustrates how the ICME could result in an asymmetric magnetic field dependence of the amplitude.
For simplicity we only consider the $y$-component of the generated effective magnetic field, since this component is responsible for the asymmetry.
(1) A sample with perpendicular magnetic anisotropy is subject to an external magnetic field $\mathbf{H}_\text{ext}$ ($-\mathbf{H}_\text{ext}$) in the $xz$-plane, pointing in the positive (negative) direction of both axes.
In equilibrium, the magnetization points along the total effective field, as indicated by the light gray arrow.
During laser pulse excitation, the ICME results in an effective magnetic field along the $y$-axis, rotating the magnetization either towards the $z$-axis or the $x$-axis, depending on the sign of $\mathbf{H}_\text{ext}$.
Additionally, the ultrafast demagnetization process leads to a reduction of the magnetization.
(2) After the laser pulse, the magnetization precesses around the total effective field that is comprised of $\mathbf{H}_\text{ext}$ and $\mathbf{H}_\text{int}$.
Depending on the sign of $\mathbf{H}_\text{ext}$, the ICME has either rotated $\mathbf{M}$ towards or away from $\mathbf{H}_\text{eff}$, resulting in different precession amplitudes.

The second coherent mechanism for laser-induced magnetization dynamics is PIMA, which leads to a step-like change in $\mathbf{H}_\text{eff}$ due to pulsed laser excitation \cite{Kalashnikova2008}.
This mechanism has been reported to arise from an optical excitation of nonequivalent lattice sites (e.g. dopants and impurities), which effectively redistributes the ions and hence changes the magnetic anisotropy \cite{Alben1972, Hansteen2006, Shen2018}.
Unlike the ICME, the PIMA mechanism is not expected to lead to an asymmetry of the magnetization precession amplitude of upon a reversal of $H_\text{ext}$, because it is present for times much longer than the period of precession and therefore acts as a constant change of the effective magnetic field \cite{Hansteen2005, Hansteen2006, Yoshimine2014, Shen2018}.

All three discussed mechanisms for inducing magnetization precession -- ICME, PIMA and the $\Delta$K mechanism -- are affected by electrostatic gating. The opto-magnetic effects can be affected through a change in e.g. the polarization-dependent refractive index and the occupation of charge states of ions and impurities.
Additionally, the $\Delta$K mechanism can be affected by the changes in charge relaxation pathways through, for example, electron-electron and electron-phonon interactions.
We find that the combination of the above mechanisms can describe quantitatively the starting phase and qualitatively the amplitude of the observed magnetic field dependence shown in Fig. \ref{fig:3} (see Supplementary Section 11).
The balance between these three mechanism affects the magnetic field dependence of the amplitude and the starting phase, increases or decreases the asymmetry in the induced precession amplitude, and changes the steepness of the starting phase versus magnetic field graph.
Therefore, since our data shows a change in these properties, we conclude that the relative strength of the mechanisms for excitation of magnetization precession are effectively controlled by electrostatic gating.

\section{Conclusions}
We envision that the electric control over the optically-induced magnetization precession amplitude demonstrated here can be applied to devices which make use of spin wave interference for signal processing \cite{Wang2017, Chumak2017, Merbouche2021}.
This should lead to an efficient electric control over the mixing of spin waves, leading to an easier on-chip implementation of combined magnonic and photonic circuits.
Even though the control over the precession frequency shown here is still modest ($\approx$10\%), we believe it can be further enhanced by the use of more effective electrostatic doping, such as using high-$\kappa$ dielectrics or ionic-liquid gating which is capable of achieving over one order of magnitude higher changes in carrier densities than the ones reported here \cite{Deng2018, Tang2022, Wang2018a, Verzhbitskiy2020, Zhuo2021}.
We note that due to the non-monotonic behavior of the magnetic anisotropy energy with changes in charge carrier density, one might expect more drastic changes on $H_\text{int}$ for larger changes in $\Delta n$.
This control over the magnetic anisotropy can then be used for the electrostatic guiding and confinement of spin waves, leading to an expansion of the field of quantum magnonics.
Finally, the presence of coherent optical excitation of magnetization dynamics we observed in CGT should also lead to a more energy-efficient optical control of magnetization \cite{Stupakiewicz2017}.
Therefore, the electric control over magnetization dynamics in CGT shown here provides the first steps towards the implementation of vdW ferromagnets in magneto-photonic devices that make use of spin waves to transport and process information.

\section{Methods}
\subsection{Sample fabrication}
The thin hBN and graphite flakes are exfoliated from bulk crystals (HQ graphene) on an oxidized silicon wafer (285 nm oxide thickness).
The CGT flakes are exfoliated in the same way in an inert (nitrogen gas) environment glove box with less than 0.5 ppm oxygen and water to prevent degradation.
The flakes are selected using optical contrast and stacked using a polycarbonate/polydimethylsiloxane stamp by a dry transfer van der Waals assembly technique \cite{Zomer2014}.
First an hBN flake (21 nm thick) is picked up, followed by the CGT flake.
Next, a thin graphite flake is picked up to make electrical contact with a corner of the CGT, and extends beyond the picked-up hBN flake.
After this, a second hBN flake (20 nm thick) is picked up and a thin graphite flake to function as the back gate electrode.
This stack is then transferred to an optically transparent fused quartz substrate finally a thin graphite flake is transferred on top the stack to function as the top gate electrode.
The device is then contacted by Ti/Au (5/50 nm) electrodes fabricated using conventional electron-beam lithography and thin metallic film deposition techniques.

\subsection{Measurement setup}
All measurements are done at 10 K under low-pressure (20 mbar) Helium gas. The sample is mounted at an angle, such that the sample normal makes an angle of 50 degrees with the external magnetic field and the laser propagation direction.

The $\sim$200 fs long laser pulses are generated by a mode-locked Ti:Sapphire oscillator (Spectra-Physics MaiTai), at a repetition rate of 80 MHz.
After a power dump, the pulses are split in an intense pump and a weaker probe pulse by a non-polarizing beam splitter.
The pump beam goes through a mechanical delay stage, allowing us to modify the time-delay between pump and probe by a change in the optical path length.
To allow for a double-modulation detection \cite{Guimaraes2018, Rojas-Lopez2023}, the pump beam goes through an optical chopper working at 2173 Hz.
The polarization of the pump is set to be horizontal (p-polarized with respect to the sample), to allow us to block the pump beam through a polarization filter at the detection stage.
The initially linearly polarized probe pulse goes through a photoelastic modulator (PEM) which modulates the polarization of the light at 50 kHz.
A non-polarizing beam splitter is used to merge the pump and probe beams on parallel paths, with a small separation between them. 
From here, they are focused onto the sample by an aspheric cold lens with a numerical aperture of 0.55.
The probe spot size (Full Width at Half Maximum) is $\sim$1.8 $\mu$m and the pump spot size is $\sim$3.4 $\mu$m, both elongated by a factor of $1/\sin(50^\circ)$ because the laser hits the sample at $50^\circ$ with respect to the sample normal.
The fluence of the pump and probe pulses are $F_\text{pump}$ = 25 $\mu$J/cm$^2$ and $F_\text{probe}$ = 5.7 $\mu$J/cm$^2$, respectively.
The transmitted light is collimated by an identical lens on the opposite side of the sample and leaves the cryostat.
The pump beam is blocked and the probe beam is sent to a detection stage consisting of a quarter wave plate, a polarization filter, and an amplified photodetector.
The quarter wave plate and the polarization filter are adjusted until they compensate for the change in polarization caused by the optical components between the PEM and the detection stage, ensuring that our signals are purely due to the rotation or ellipticity of the probe polarization induced by our samples.
The first and second harmonic of the signal (50 or 100 kHz) obtained at the photodetector are then proportional to the change in ellipticity and rotation due to the Faraday effect of the sample.
For static magneto-optic Faraday effect measurements we have blocked the pump beam before reaching the sample.

\subsection{Calculating $\Delta n$ and $\Delta D$ from the gate voltages}
The gate-induced change in charge carrier density ($\Delta n$) and displacement field ($\Delta D$) at the CGT are calculated from the applied gate voltages using a parallel plate capacitor model.
The displacement field generated by the top ($D_\text{t}$) and back ($D_\text{b}$) gates is given by $D_i = \varepsilon_\text{hBN} E_i = \frac{1}{2} \sigma_{\text{free},i}$, where $i$ denotes $t$ or $b$, $\varepsilon_\text{hBN}$ = $3.8\varepsilon_0$ is the hBN dielectric constant \cite{Laturia2018} with $\varepsilon_0$ the vacuum permittivity, and $\sigma_\text{free}$ the free charge per unit area. The applied top and back gate voltages are related to $\sigma_\text{free}$ by $V_\text{i} = -\int D_\text{i}/\varepsilon \, \mathrm{d}z$.
This equation, combined with the condition of charge neutrality, gives the following 3 relations:
\begin{align*}
V_t / d_t &= \frac{\sigma_t - \sigma_\text{CGT} - \sigma_b}{2\varepsilon_\text{hBN}},\\
V_b / d_b &= \frac{\sigma_b - \sigma_\text{CGT} - \sigma_t}{2\varepsilon_\text{hBN}},\\
0 &= \sigma_t + \sigma_b + \sigma_\text{CGT},
\end{align*}
where $d_\text{t,b}$ denotes the thickness of the top (21 nm) and bottom (20 nm) hBN flakes,  and $\sigma_i$ the free charge per unit area in the top gate ($t$), back gate ($b$), and the CGT flake. Solving this set of equations yields:
\begin{align*}
\sigma_t &= \varepsilon_\text{hBN} V_t / d_t ,\\
\sigma_b &= \varepsilon_\text{hBN} V_b / d_b ,\\
\Delta n = \sigma_\text{CGT}/e &= -\frac{\varepsilon_\text{hBN}}{e} \left(\frac{V_t}{d_t} + \frac{V_b}{d_b}\right),\\
\end{align*}
where $e$ is the positive elementary charge.
Note that for positive gate voltages, a negative charge carrier density is induced in the CGT.
For the gate-induced change in the displacement field at the CGT layer, we get:
\begin{align*}
\Delta D = (\sigma_b - \sigma_t)/2 = -\varepsilon_\text{hBN} (V_t / d_t - V_b / d_b) / 2
\end{align*}
Filling in the values for the thickness of the hBN flakes and dielectric constant of hBN gives $\Delta D/\varepsilon_0$ and $\Delta n$ at the CGT:
\begin{align*}
    \Delta n &= -\left(1.00 V_t + 1.05 V_b\right) \times 10^{12} V^{-1}\text{cm}^{-2}\\
    D/\varepsilon_0 &= -\left(0.090 V_t - 0.095 V_b\right) \text{nm}^{-1}.
\end{align*}
Throughout the main text we use $\Delta D/\varepsilon_0$ instead of $\Delta D$ for easier comparison of our values of the gate-induced change in the displacement field with values mentioned in other works.
Note that we use the conversion factor $\varepsilon_0$ and not the permittivity of CGT.
Therefore, the values for the $\Delta D$ that we report are the equivalent electric field values in \emph{vacuum}, not in CGT.

\subsection{Windowed Fourier transform}
The RMS power spectra of the TRFE oscillations shown in Fig. \ref{fig:1}e are calculated from the TRFE measurements using a windowed Fourier transform.
The type of window used for this calculation if the Hamming window, which extends from $\Delta t = 0$ up to the last data point.
The RMS power spectrum ($P_\text{RMS}(f)$) of the TRFE oscillations is calculated as
\begin{align*}
    P_\text{RMS}(f) = \Bigg(\sum_{\Delta t > 0} &\left[ W_\text{Ham}(\Delta t) y(\Delta t) \sin(2\pi f \Delta t)\right]^2 +\\
    &\left[ W_\text{Ham}(\Delta t) y(\Delta t) \cos(2\pi f \Delta t)\right]^2 \Bigg)^{1/2},
\end{align*}
where $W_\text{Ham}$ is the Hamming window, y the data points of the TRFE measurements, and $f$ the frequency.

\subsection{Determining the $g$-factor and $H_\text{int}$}
The Land\'{e} $g$-factor and $H_\text{int}$ can be extracted by fitting the magnetic field dependence of the precession frequencies with Eq. (\ref{eq:frequency_simple}).
The values of $g$ and $H_\text{int}$ we obtained from the fit were, in most cases, strongly correlated.
Therefore, we first determined $g$ by fitting the data for $\mu_0 H_\text{ext} \geq 125$ mT, since $g$ is most sensitive to the slope at high fields.
This yields $g$ = 1.89 $\pm$ 0.01.
I we further allow for an additional uncertainty in the mounting angle of the sample, the g-factor can change by $\sim 0.1$.
Then we determine $H_\text{int}$ by fitting Eq. (\ref{eq:frequency_simple}) for all remaining measurements fixing $g$ = 1.89.
We note that the values for $H_\text{int}$ do depend on the exact value of $g$, but the modulation due to electrostatic gating is not affected, as is shown in Supplementary Section 4.

\subsection{Extracting the magnetization precession parameters from the TRFE measurements}
We extract the amplitude, frequency, and starting phase of the oscillations in the TRFE measurements by fitting the data for $\Delta t > 26$ ps with the phenomenological formula \cite{Mizukami2010, Zhang2020}
\begin{align}
    y = &y_0 + ae^{-\Delta t / \tau_\text{osc}}\cos\left(2\pi f \Delta t - \phi_0\right) \nonumber\\
    &+ A_\text{l} e^{-\Delta t / \tau_\text{l}}+ A_\text{s} e^{-\Delta t / \tau_\text{s}}. \label{eq:trfr_fit}
\end{align}

This formula describes a phase shifted sinusoid on top a double exponential background.
The background captures the demagnetization and remagnetization of the CGT, while the sinusoid describes the magnetization precession.

\section{Data availability}
The raw data and the data underlying the figures in the main text are publicly available through the data repository Zenodo at https://doi.org/10.5281/zenodo.8321758.

\section{Acknowledgments}
We thank Bart J. van Wees for critically reading the manuscript and providing valuable feedback, and we thank J. G. Holstein, H. Adema, H. de
Vries, A. Joshua and F. H. van der Velde for their technical support.
This work was supported by the Dutch Research Council (NWO) through grants STU.019.014 and OCENW.XL21.XL21.058, the Zernike Institute for Advanced Materials, the research program “Materials for the Quantum Age” (QuMat, registration number 024.005.006), which is part of the Gravitation program financed by the Dutch Ministry of Education, Culture and Science (OCW), and the European Union (ERC, 2D-OPTOSPIN, 101076932). Views and opinions expressed are however those of the author(s) only and do not necessarily reflect those of the European Union or the European Research Council. Neither the European Union nor the granting authority can be held responsible for them.
The device fabrication and nanocharacterization were performed using Zernike NanoLabNL facilities.

\section{Author information}
M.H.D.G. conceived and supervised the research.
F.H. designed and fabricated the samples, performed the measurements, analyzed the data, and calculated the effect of coherent excitations on the magnetization precession under M.H.D.G. supervision.
F.H. and R.R.R.L built and tested the measurement setup.
F.H., M.H.D.G, and B.K discussed the data and provided the interpretation of the results.
F.H. and M.H.D.G co-wrote the manuscript with input from all authors.

\section{Ethics declaration}
\subsection{Competing interests}
The authors declare no competing interests.

%
\end{document}


\title{Supplementary Information for 'Electric control of optically-induced magnetization dynamics in a van der Waals ferromagnetic semiconductor'}

\author{Freddie Hendriks}
\affiliation{Zernike Institute for Advanced Materials, University of Groningen, The Netherlands}

\author{Rafael R. Rojas-Lopez}
\affiliation{Zernike Institute for Advanced Materials, University of Groningen, The Netherlands}
\affiliation{Departamento de F\'{i}sica, Universidade Federal de Minas Gerais, Brazil}

\author{Bert Koopmans}
\affiliation{Department of Applied Physics, Eindhoven University of Technology, The Netherlands}

\author{Marcos H. D. Guimar\~{a}es}
\affiliation{Zernike Institute for Advanced Materials, University of Groningen, The Netherlands}
\email{m.h.guimaraes@rug.nl}

\date{22 September 2023}
\maketitle

\section{Atomic force microscopy}
\begin{figure*}[htbp]
	\centering
	\includegraphics[width=\textwidth]{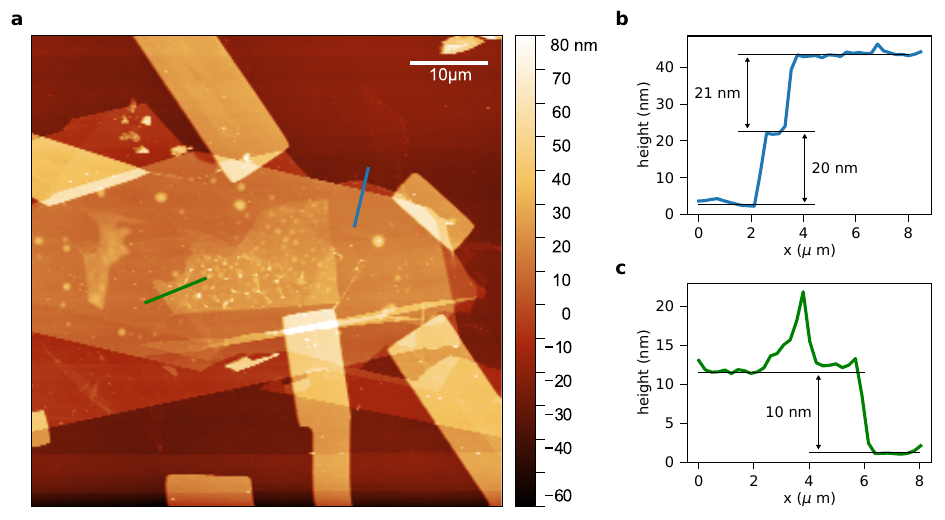}
	\caption{\textbf{Atomic force microscopy scan.} \textbf{a}, AFM height image. Height profiles taken along the blue and green line are used to extract the thickness of the top and bottom hBN, and of the thickness of the CGT respectively.
    \textbf{b}, Height profile taken along the blue line to extract the thickness of the top (21 nm) and bottom (20 nm) hBN flakes. 
    \textbf{c}, Height profile taken along the green line to extract the thickness of the CGT flake (10 nm).
    }
	\label{si_fig:AFM}
\end{figure*}

The thicknesses of the flakes of the sample were measured using atomic force microscopy (AFM).
The height map is shown in Fig. \ref{si_fig:AFM}a.
The thickness of the hexagonal boron nitride (hBN) flakes are determined from the line trace along the blue line, shown in Fig. \ref{si_fig:AFM}b, yielding 20 nm for the bottom and 21 nm for the top flake respectively.
The Cr$_2$Ge$_2$Te$_6$ (CGT) thickness was extracted from the line profile along the green line, shown in Fig. \ref{si_fig:AFM}c, yielding 10 nm.

\FloatBarrier
\section{Finding optimal parameters for the Faraday effect}
The sensitivity of the Faraday effect on changes in the magnetization depends on the wavelength.
Additionally, the change in rotation and ellipticity of the light caused by the Faraday effect are generally different.
To find the optimal wavelength and polarization mode, we probe the magnetization using the Faraday effect while sweeping the magnetic field.
The change in polarization when the magnetic field reverses the magnetization direction of the CGT indicates the sensitivity of the Faraday effect.
We determine this sensitivity for various values of the wavelength, ranging from 830 nm to 940 nm, measuring change in rotation and ellipticity simultaneously.
We found that the ellipticity at 870 nm had the highest sensitivity, and therefore we use this wavelength and polarization mode for the time-resolved Faraday ellipticity (TRFE) measurements.

\FloatBarrier
\section{Gate voltage dependence of magnetization curves measured by static Faraday ellipticity}
\begin{figure*}[htbp]
	\centering
	\includegraphics[width=\textwidth]{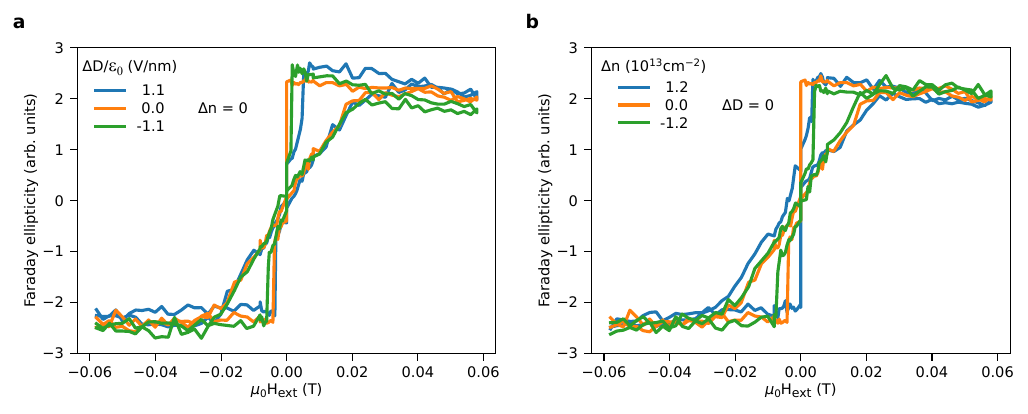}
	\caption{\textbf{Gate dependence of the magnetization curves measured by static Faraday ellipticity.}
    \textbf{a}, Effect of $\Delta D$ on the magnetization curves for $\Delta n = 0$.
    \textbf{b}, Effect of $\Delta n$ on the magnetization curves for $\Delta D = 0$.
    }
	\label{si_fig:magnetization_curves}
\end{figure*}

We measured the gate dependence of the magnetization curves of CGT by means of the Faraday effect using a pulsed laser with a fluence of 5.7 $\mu$J/cm$^2$ at a wavelength of 870 nm.
We measured the change in ellipticity of the light, as described in the Methods.
The gate dependence of the magnetization curves for various values of $\Delta n$ and $\Delta D$ are presented in Fig. \ref{si_fig:magnetization_curves}.

\begin{figure*}[htbp]
	\centering
	\includegraphics[width=\textwidth]{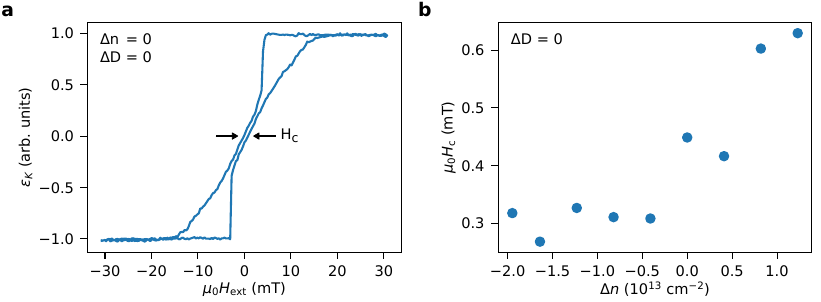}
	\caption{\textbf{Coercive field from static MOKE measurements}
    \textbf{a}, Magnetization curve measured by static MOKE at $\Delta n = \Delta D = 0$. The coercive field is half of the horizontal shift of the trace and retrace near zero.
    \textbf{b}, Effect of $\Delta n$ on the coercive field at $\Delta D = 0$. 
    }
	\label{si_fig:moke_coercivity}
\end{figure*}

In addition, we measured the magnetization curves at 920 nm using the polar magneto-optic Kerr effect (MOKE) in a different experimental setup as a function of $\Delta n$ at $\Delta D = 0$.
From this data we extracted how the coercive field changes with $\Delta n$, which is shown in Fig. \ref{si_fig:moke_coercivity}, to check if it is similar to the $\Delta n$ dependence of $H_\text{int}$.
We extract the coercive field by individually fitting the trace and retrace of the magnetization curves with the formulas
\begin{align}
    y_\text{trace} &=
    \begin{cases}
        A \tanh\left(\frac{x - x_{0,t}}{w}\right) + ax + b, & \text{if } x > x_\text{s,t}\\
        A + ax + b, & \text{otherwise}
    \end{cases}\\
    y_\text{retrace} &= 
    \begin{cases}
        A \tanh\left(\frac{x - x_{0, r}}{w}\right) + ax + b, & \text{if } x < x_\text{s,r}\\
        A + ax + b, & \text{otherwise}
    \end{cases}
\end{align}

where $A$ is half of the difference in MOKE ellipticity for large positive and negative fields, $x_0$ the horizontal shift, $w$ the width of the $\tanh$ (which is proportional to the saturation field), and $x_s$ the position of the sharp switch close to zero field that indicates the transition from a single domain to multi domain state \cite{Lohmann2019}.
The parameters $a$ and $b$ are respectively the slope and offset of the background.
The coercive field is generally not fitted well.
To get a better estimate for the coercive field, we subtract the background, and separately fit a straight line, with equal slope, through the trace and retrace  where the signal is close to zero and approximately linear (we used the signal value $<$ 0.002).
The coercive field is then calculated as half of the horizontal separation of these two lines.
The graph of $\mu_0 H_c$ versus $\Delta n$ presented in Fig. \ref{si_fig:moke_coercivity} is very similar to the graph of $\mu_0 H_\text{int}$ versus $\Delta n$ at $\Delta D = 0$ presented in Fig. 2d in the main text.
This means that the $\Delta n$ dependence of $H_\text{int}$ obtained from TRFE measurements of the magnetization precession is consistent with the $\Delta n$ dependence of $H_c$ obtained from static MOKE mearements.

\FloatBarrier
\section{$\theta_H$ dependence of $g$ and $H_\text{int}$}
\begin{figure*}[htbp]
	\centering
	\includegraphics[width=\textwidth]{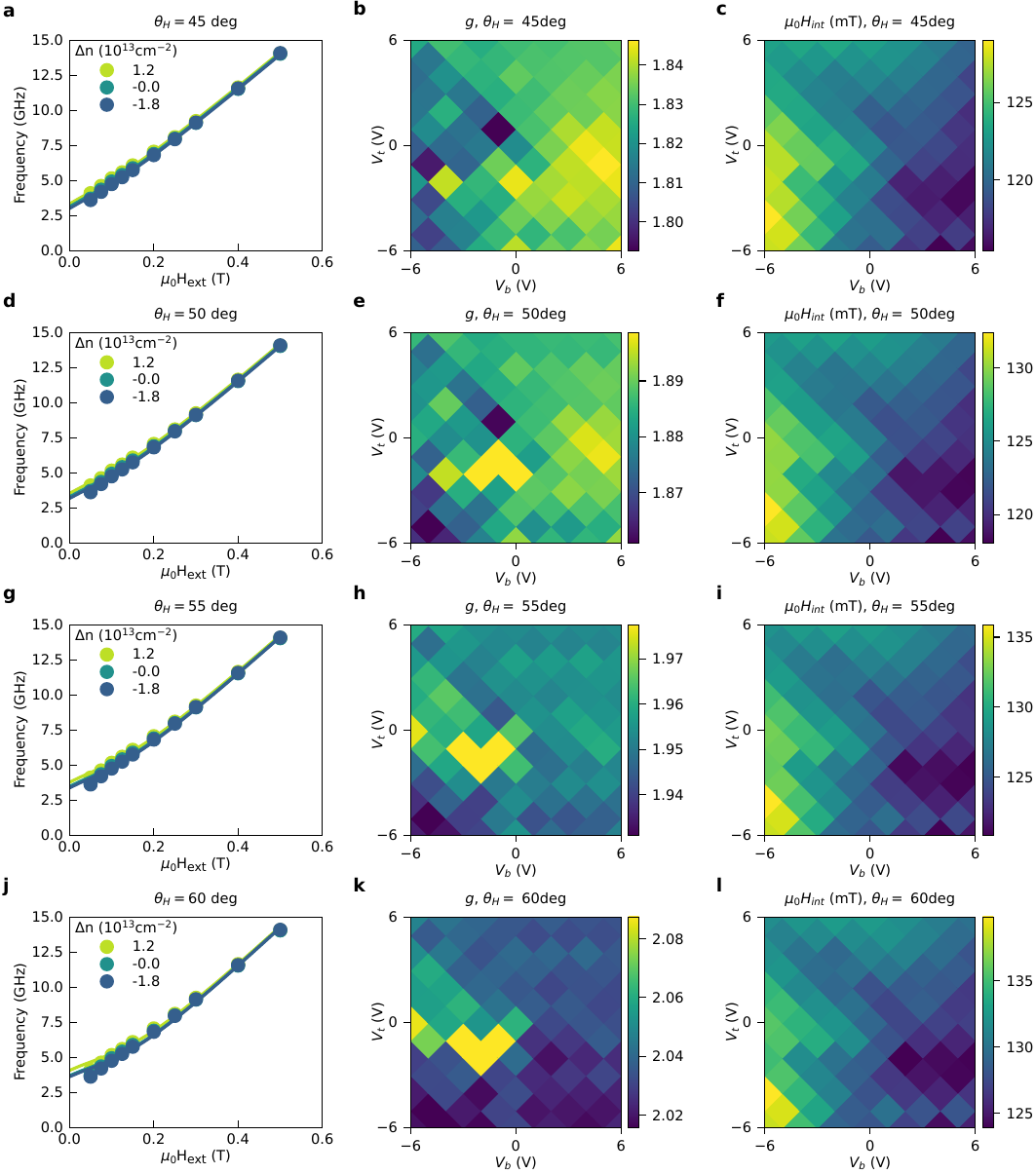}
	\caption{\textbf{Effect of the sample mounting angle on $g$ and $H_\text{int}$.}
    Fit of precession frequency versus $H_\text{ext}$ (\textbf{a}) data to obtain $g$ (\textbf{b}) and $H_\text{int}$ (\textbf{c} for) $\theta_\text{H} = 45$ deg.
    Results for $\theta_H = 50, 55$ and $60$ degrees are presented in respectively \textbf{d-f}, \textbf{g-h} and \textbf{j-l}.
    }
	\label{si_fig:theta_h_dependence}
\end{figure*}

We extract the $g$-factor and the internal effective field ($H_\text{int}$) from the external magnetic field ($H_\text{ext}$) dependence of the magnetization precession frequency ($f$), as explained in the main text and the Methods.
To obtain $g$ and $H_\text{int}$, we need to fix the value of the angle $\theta_H$ between $H_\text{ext}$ and the sample normal.
Fig. \ref{si_fig:theta_h_dependence} shows the gate dependence of $g$ and $H_\text{int}$ for different values of $\theta_\text{H}$, ranging from 45 degrees to 60 degrees.
Panels a, d, g, and j show that for $\theta_\text{H} = 45^\circ$ and $50^\circ$ the data is fitted best.
Since $50^\circ$ is more likely based on how we mounted the sample in the cryostat, we have chosen $\theta_\text{H} = 50^\circ$ for our analysis, despite the observation that $\theta_H = 60^\circ$ results in a $g$-factor that is more in line with values reported in literature.
From the results presented in Fig. \ref{si_fig:theta_h_dependence}, we can conclude that, apart from an offset of the mean value, the gate dependence of $g$ and $H_\text{int}$ hardly changes with $\theta_n$, even if we choose $\theta_\text{H}$ such that we obtain a $g$ factor that agrees with literature values.
Therefore, the conclusions in the main on the gate dependence of the $H_\text{int}$ do not depend on the exact value of $\theta_H$, as long as these values are reasonable.

\FloatBarrier
\section{Pump power dependence of precession frequency}
\begin{figure*}[htbp]
	\centering
	\includegraphics[width=\textwidth]{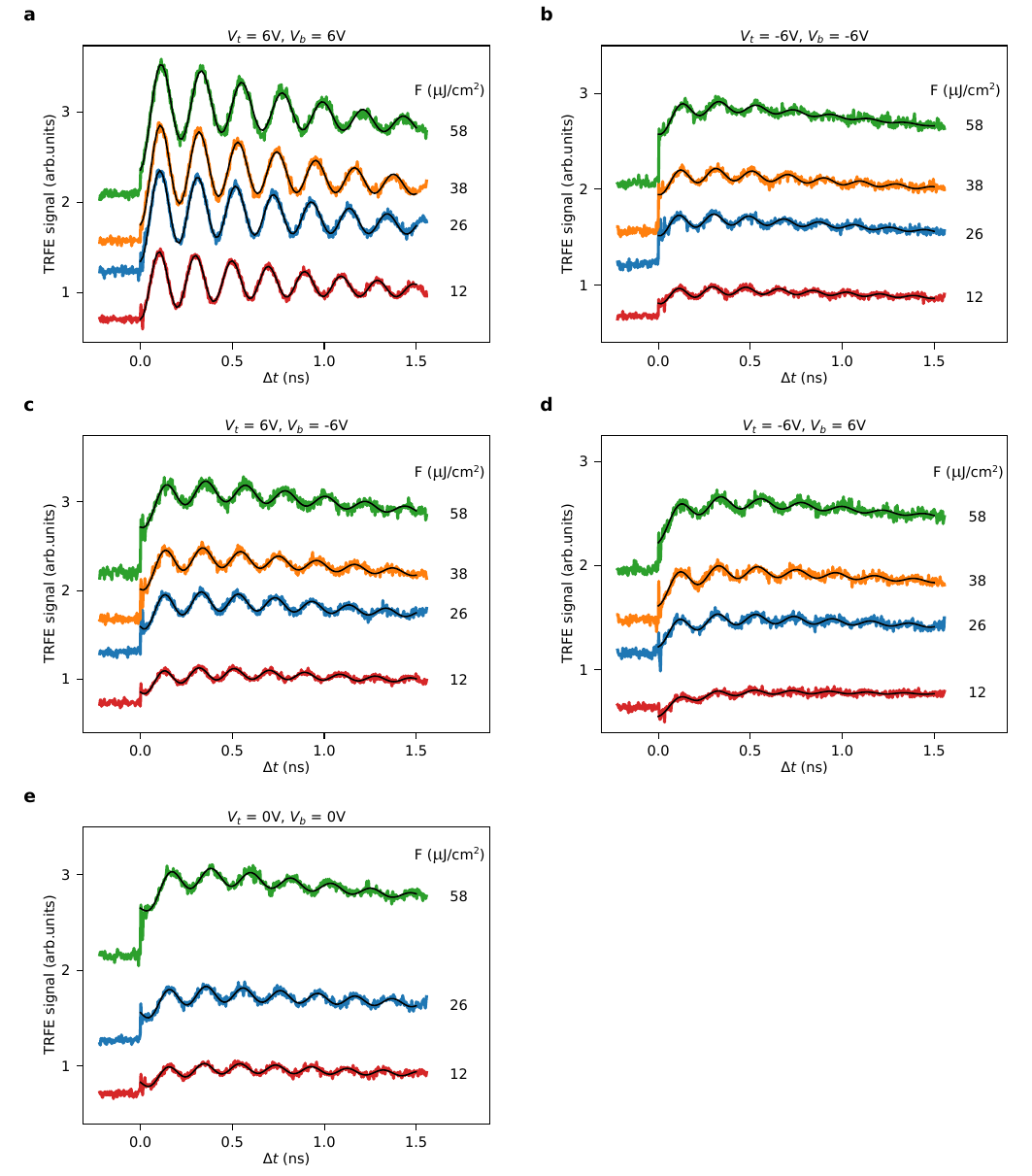}
	\caption{\textbf{Pump fluence dependence of the TRFE measurements.} Dependence of the TRFE signal on the pump fluence, measured at $\mu_0 H_\text{ext} = 100$ mT, for various values of $\text{V}_{t}$ and $\text{V}_{b}$. The solid lines indicate the best fit of Eq. (2) in the Methods, which is used to extract the magnetization precession frequency.
    }
	\label{si_fig:TRFE_pumppower}
\end{figure*}

\begin{figure*}[htbp]
	\centering
	\includegraphics[]{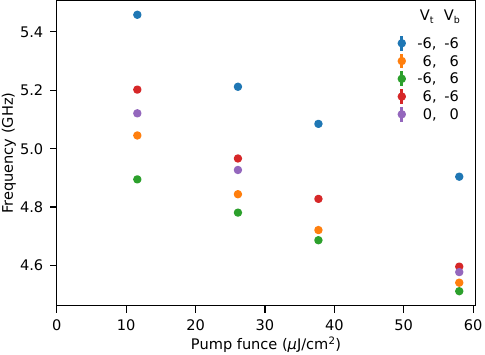}
	\caption{\textbf{Pump fluence dependence of the precession frequency.} Dependence of the precession frequency on the pump fluence, measured at $\mu_0 H_\text{ext} = 100$ mT, extracted from Fig. \ref{si_fig:TRFE_pumppower}. All the extreme gate voltage combinations used for the gate voltage scans of the magnetization dynamics (Figs. 2a, 3a, and 3b of the main text) are shown.
    }
	\label{si_fig:freq_pumppower}
\end{figure*}

We found that the frequency of the magnetization precession depends on the fluence of the pump pulse. 
For the data presented in the main text, the pump fluence was measured to be 25 $\mu$J/cm$^2$.
Fig. \ref{si_fig:TRFE_pumppower} shows how the TRFE signal is affected by different values of the pump fluence.
The frequency for each of these measurements is extracted by fitting the data with Eq. (2) of the Methods, and is plotted in Fig. \ref{si_fig:freq_pumppower}.
There it can be seen than the magnetization precession frequency decreases with pump fluence for all the extreme gate voltage combinations.
The precise rate at which the frequency depends on the pump power depends on the applied gate voltages.
This can be due the photo-induced magnetic anisotropy effect (PIMA) since an increase in pump fluence would result on an increase on the increase (or decrease) of the population of the dopant or impurity states.
Since the PIMA effect depends on the population of these states, an increase in the pump fluence should result on an increase on the effective field generated by it.
The change in the observe frequency should then depend on the specific sign (with respect to the magnetic anisotropy and applied magnetic field) of the PIMA.
Additionally, a decrease in the saturation magnetization ($M_\text{s}$) and magnetocrystalline anisotropy caused by heating by the laser could offer another possible explanation.
This does affect the values of the Land\'{e} g-factor and the effective internal field ($H_\text{ext}$).
Since we used the same pump fluence for all measurements, it does not affect the general trends we observe for the gate voltage dependence of the magnetization precession frequency, starting phase, and amplitude shown in this work.

\FloatBarrier
\section{Damping of magnetization precession}
The damping time of the magnetization precession ($\tau_\text{osc}$) is an important parameter to describe the magnetization dynamics.
As mentioned in the main text, the precession frequency we measured is well described by the Landau-Lifshitz-Gilbert (LLG) equation in the limit of low damping.
In the LLG equation, the effective damping is included through the phenomenological dimensionless parameter $\alpha_\text{eff}$.
This damping describes how the magnetization relaxes to its equilibrium state after excitation.
The effective damping parameter $\alpha_\text{eff}$ can be decomposed into a frequency-independent intrinsic damping $\alpha$, and a frequency-dependent extrinsic damping $\alpha_\text{ext}$, resulting in $\alpha_\text{eff} = \alpha + \alpha_\text{ext}$.
For our measurements we assume that the main source of extrinsic damping is spatial fluctuations in $H_\text{int}$.
This contributes to the damping via inhomogeneous broadening of the precession frequency, and is quantified by $\alpha_\text{ext} = \sqrt{ 2\ln(2)}\left| \text{d}\omega/dH_\text{int} \right| \Delta H_\text{int} / \left[\frac{g\mu_B \mu_0}{\hbar} (2\left|\mathbf{H}_\text{ext} \right| - H_\text{int} \sin^2(\theta_M))\right]$ \cite{Zhang2020}.
Here, $g$ is the Land\'{e} g-factor, $\mu_B$ the Bohr magneton, $\mu_0$ the vacuum permeability, $\mathbf{H}_\text{ext}$ the external magnetic field, $\theta_M$ the angle between the magnetization and the sample normal, $\omega$ the angular frequency of the precession, and $H_\text{int} = 2K_u / (\mu_0 M_\text{s}) - M_\text{s}$ with $K_u$ the uniaxial magnetic anisotropy energy and $M_\text{s}$ the saturation magnetization.
Note that in \cite{Mizukami2010} $\left|\mathbf{H}_\text{eff}\right|$ is expressed as $H_\text{ext} \cos(\theta - \theta_M) + H_\text{int} \cos^2(\theta_M)$.
The angle $\theta_M$ is calculated by minimizing the magnetic energy in the presence of an external field, perpendicular magnetic anisotropy, and shape anisotropy \cite{Zhang2020}.
The derivative $\left| \text{d}\omega/dH_\text{int} \right|$ is calculated numerically using Eq. (1) of the main text.
For large external magnetic fields, $\Delta H_\text{int}$ becomes less relevant, and $\alpha_\text{eff} \approx \alpha$.


The expression for the magnetization precession damping time ($\tau_\text{osc}$) obtained from the ferromagnetic resonance (FMR) mode of the LLG equation for small values of $\alpha_\text{eff}$ is given by \cite{Mizukami2010, Zhang2020}
\begin{align}
    \tau_\text{osc} = 2\hbar / \left[g \mu_B \mu_0 \alpha_\text{eff} \left(2\left|\mathbf{H}_\text{ext} \right| - H_\text{int} \sin^2(\theta_M)\right)\right].
    \label{si_eq:tau_simple}
\end{align}

\begin{figure*}[htbp]
	\centering
	\includegraphics[]{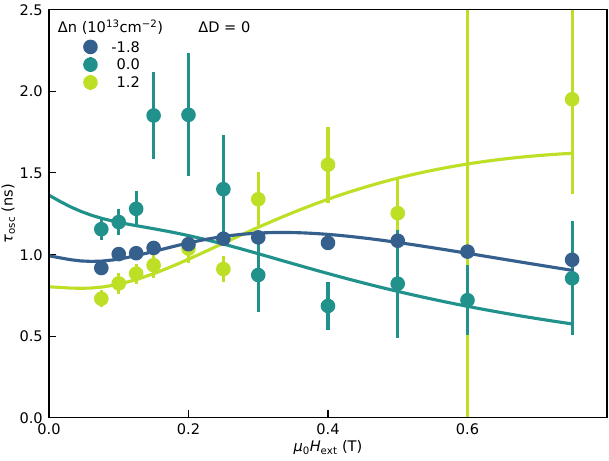}
	\caption{\textbf{Gate dependence of precession decay time.} 
    Gate dependence of the relation between decay time of the magnetization precession and $H_\text{ext}$. The circles indicate decay times extracted from the TRFE measurements. The solid lines correspond to the best fit of Eq. (\ref{si_eq:tau_simple})
    }
	\label{si_fig:damping}
\end{figure*}

To obtain the damping parameters $\alpha$ and $\Delta H_\text{int}$, this equation is fitted to the values of $\tau_\text{osc}$ extracted from the TRFE measurements at various values of $\Delta n$ with $\Delta D = 0$, which is presented in Fig. \ref{si_fig:damping}.
We used the values $g = 1.886$ and $\mu_0 H_\text{int} = 134$, $125$ and $120$ mT for respectively $\Delta n = 1.2$, $0$ and $-1.8 \times10^{13}$cm$^{-2}$, which are obtained from fitting the precession frequency versus $H_\text{ext}$ data using Eq. (1) of the main text.
The fit results for the damping parameters are summarized in Tab. \ref{si_tab:damping_results}.
As can be seen in Fig. \ref{si_fig:damping}, the data for $\Delta n = -1.8\times10^{13}$cm$^{-2}$ is fitted well by Eq. (\ref{si_eq:tau_simple}).
The data for $\Delta n = 1.2\times10^{13}$cm$^{-2}$ is fitted reasonably well, but for $\Delta n = 0$ the fit is quite poor, even when taking the errors in $\tau_\text{osc}$ into account.

A bad fit likely results from the large spread of the measurements points at high external magnetic fields, and (when present) from the peak in $\tau_\text{osc}$ around $H_\text{ext} = 200$ mT.
This peak, which starts around $H_\text{int} \approx 125$ mT, consistently shows up for other small values of $\Delta n$ and $\Delta D$ as well.
A possible explanation for this is that there are more factors affecting the extrinsic damping than $\Delta H_\text{int}$ alone.
The large spread and errors in $\tau_\text{osc}$ for $\Delta n = 0$ and $1.2 \times 10^{13}$cm$^{-2}$ are caused by a combination of two factors.
The first factor is the low precession amplitude, which results in a lower signal to noise ratio for all fitting parameters related to the amplitude of the precession.
This can be improved by inducing precession with a larger amplitude (e.g. higher pump fluence), or by reducing the noise in the system (e.g. higher probe fluence, or better blocking of pump before the detection stage).
The second factor is the relatively short maximum pump-probe delay ($\Delta t_\text{max}$) that we can achieve in our experimental setup, which is about 1.5 ns.
For a good estimate of the magnetization precession damping time, $\Delta t_\text{max} \gg \tau_\text{osc}$.
In our case however, $\Delta t_\text{max} \approx \tau_\text{osc}$.

Using all three fit results, we obtain an intrinsic Gilbert damping of $\alpha \approx6\times10^{-3}$, and a spread in the internal effective field of $\Delta H_\text{ext} \approx 9$ mT.
Both of these values are in line with values found in literature for thin CGT flakes \cite{Zhang2020, Zollitsch2022}.

\begin{table}[htbp]
\begin{tabular}{c|lc}
$\Delta n$ $(10^{13}$cm$^{-2})$ & \multicolumn{1}{c}{$\alpha$ $(10^{-3})$} & $\mu_0\Delta H_\text{int}$ (mT) \\ \hline
-1.8                            & $6.5 \pm 0.3$                              & $9.0 \pm 0.2$                   \\
0                               & $12 \pm 3$                               & $5 \pm 1$                       \\
1.2                             & $2 \pm 1$                                & $12.6 \pm 0.7$                  \\ \hline
\end{tabular}
\caption{\textbf{Damping parameters.}
Fit results of the damping of the magnetization precession for three values of $\Delta n$ with $\Delta D = 0$. The error in these values is the statistical error obtained from the least squares fitting procedure.}
\label{si_tab:damping_results}
\end{table}
\FloatBarrier
\section{Detailed discussion on other possible mechanisms of the TRFE amplitude modulation caused by electrostatic gating}
\label{si_sec:trfe_ampl_mod}
The change in amplitude of the oscillations in the TRFE signal with gating can have multiple causes.
The most obvious one is a change in the magnetization precession amplitude.
It could however also be caused by a change in the strength of the Faraday effect.
In this case, the TRFE curves for different values of the gate voltages (measured at the same external magnetic field) would only differ by scale factor, meaning that the curves can be mapped onto each other by simply scaling them along the vertical axis.
This is clearly not happening in the measurements shown in Fig. 1a in the main text and Fig. \ref{si_fig:trfe_negative_fields}, and therefore the change in amplitude is not caused by a pure change in the strength of the Faraday effect.
Gate dependent magnetization curve measurements performed in the same setup, shown in Fig. \ref{si_fig:magnetization_curves}, confirm that the strength of the Faraday effect is not significantly changed by electrostatic gating. 

Another possibility would be that the equilibrium angle of $\mathbf{M}$ changes due to a change in the induced charge carrier density ($\Delta n$).
This then would change the projection of the precession plane onto the propagation direction of the laser, and could change the TRFE amplitude without changing the magnetization precession amplitude.
The canting would be due to a change in the effective internal field ($H_\text{int}$).
However, our measurements show that the change in $H_\text{eff}$ is less than 15\%, which is not enough to explain the large increase in amplitude of the TRFE oscillations.

\FloatBarrier
\section{Complete data sets for the gate voltage dependence of the magnetization precession parameters}
This section displays the complete data sets of the gate voltage dependence of the magnetization precession amplitude ($A$), starting phase ($\phi_0$), and frequency ($f$).

\begin{figure*}[htbp]
	\centering
	\includegraphics[width=\textwidth]{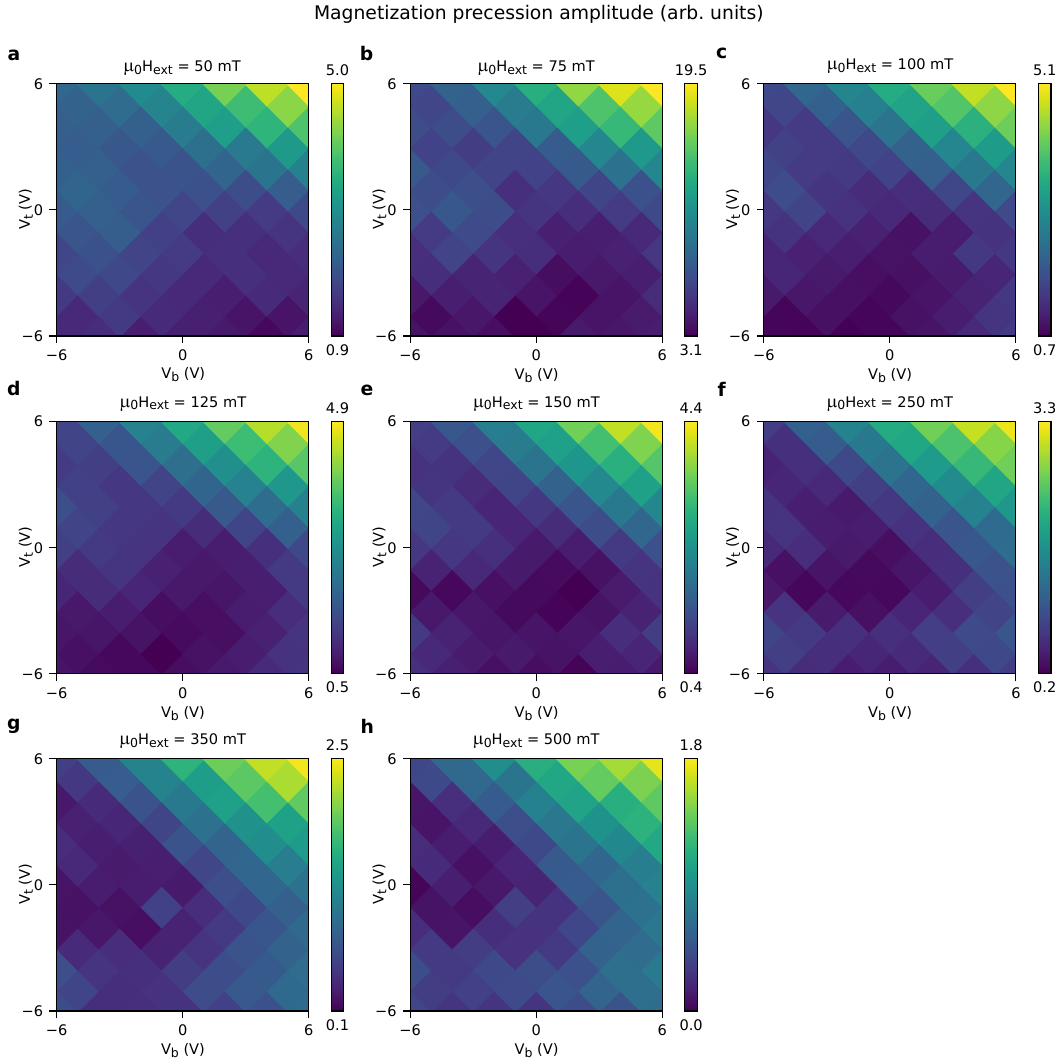}
	\caption{\textbf{Gate voltage dependence of the magnetization precession amplitude.}
    \textbf{a-h}, The complete data set of the gate voltage dependence of the magnetization precession amplitude extracted from the TRFE measurements, for all values of the magnetic fields used during the measurements. The color scale has a different range for each plot. The scale bar limits indicate the minimum and maximum amplitude, all in the same arbitrary units, for $\mu_0 H_\text{ext}$ equal to 50 mT (\textbf{a}), 75 mT (\textbf{b}), 100 mT (\textbf{c}), 125 mT (\textbf{d}), 150 mT (\textbf{e}), 250 mT (\textbf{f}), 350 mT (\textbf{g}), and 500 mT (\textbf{h}).
    }
	\label{si_fig:ampl_gate}
\end{figure*}

\begin{figure*}[htbp]
	\centering
	\includegraphics[width=\textwidth]{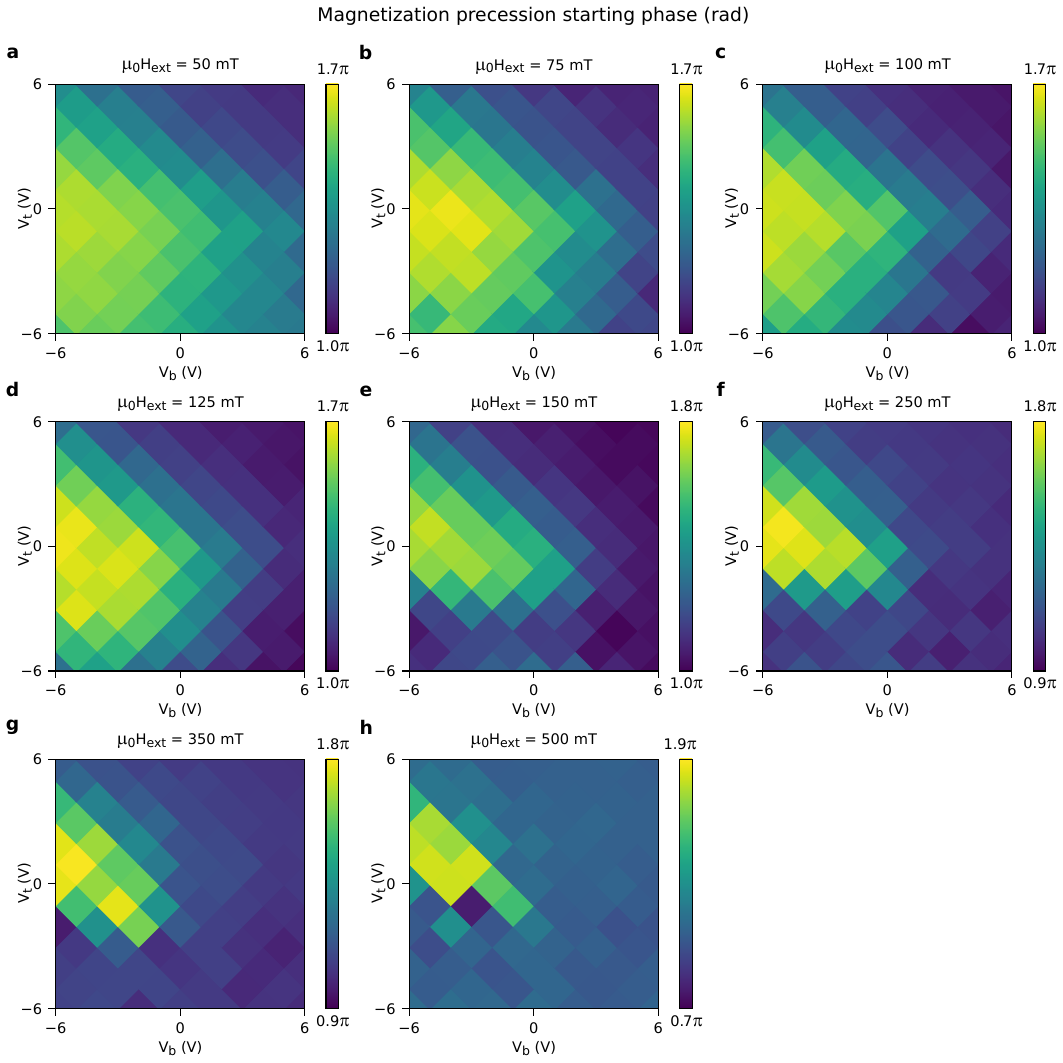}
	\caption{\textbf{Gate voltage dependence of the magnetization precession starting phase.}
    \textbf{a-h}, The complete data set of the gate voltage dependence of the magnetization precession starting phase extracted from the TRFE measurements, for all values of the magnetic fields used during the measurements. The color scale has a different range for each plot. The limits indicate the minimum and maximum phase, in radians, for $\mu_0 H_\text{ext}$ equal to 50 mT (\textbf{a}), 75 mT (\textbf{b}), 100 mT (\textbf{c}), 125 mT (\textbf{d}), 150 mT (\textbf{e}), 250 mT (\textbf{f}), 350 mT (\textbf{g}), and 500 mT (\textbf{h}).
    }
	\label{si_fig:phase_gate}
\end{figure*}

\begin{figure*}[htbp]
	\centering
	\includegraphics[width=\textwidth]{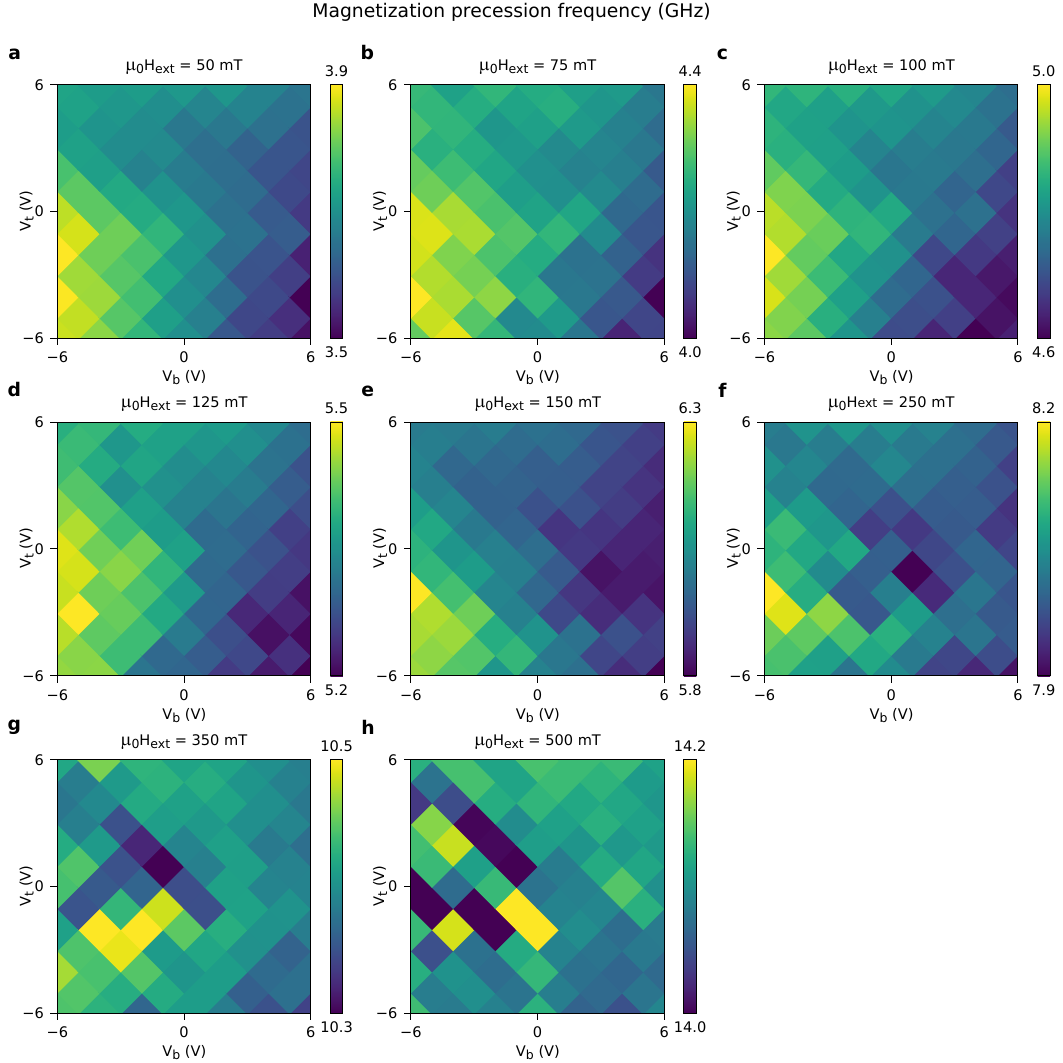}
	\caption{\textbf{Gate voltage dependence of the magnetization precession frequency for all magnetic fields.}
    \textbf{a-h}, The complete data set of the gate voltage dependence of the magnetization precession frequency extracted from the TRFE measurements, for all values of the magnetic fields used during the measurements. The color scale has a different range for each plot. The limits indicate the minimum and maximum frequency, in GHz, for $\mu_0 H_\text{ext}$ equal to 50 mT (\textbf{a}), 75 mT (\textbf{b}), 100 mT (\textbf{c}), 125 mT (\textbf{d}), 150 mT (\textbf{e}), 250 mT (\textbf{f}), 350 mT (\textbf{g}), and 500 mT (\textbf{h}).
    }
	\label{si_fig:freq_gate}
\end{figure*}

\FloatBarrier
\section{Magnetization precession frequency for negative magnetic fields}
\begin{figure*}[htbp]
	\centering
	\includegraphics[width=\textwidth]{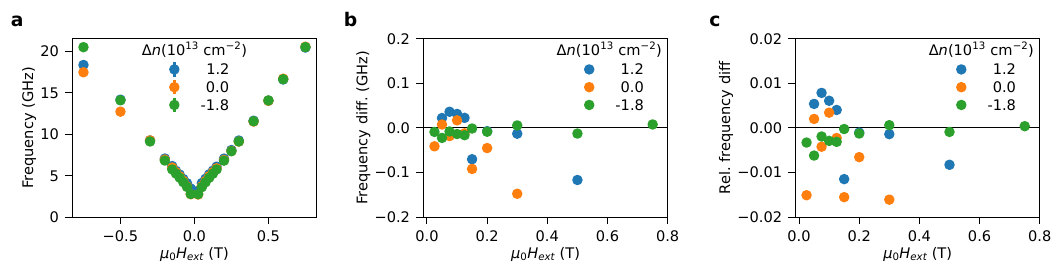}
	\caption{\textbf{$H_\text{ext}$ symmetry of the magnetization precession frequency}.
    \textbf{a}, Magnetization precession frequency for positive and negative external magnetic fields.
    \textbf{b}, Difference in precession frequency for positive and negative external magnetic fields.
    \textbf{c}, Relative difference in precession frequency for positive and negative external magnetic fields.
    All data presented has $\Delta D = 0$.
    }
	\label{si_fig:freq_field-symmetry}
\end{figure*}

\begin{figure*}[htbp]
	\centering
	\includegraphics[width=\textwidth]{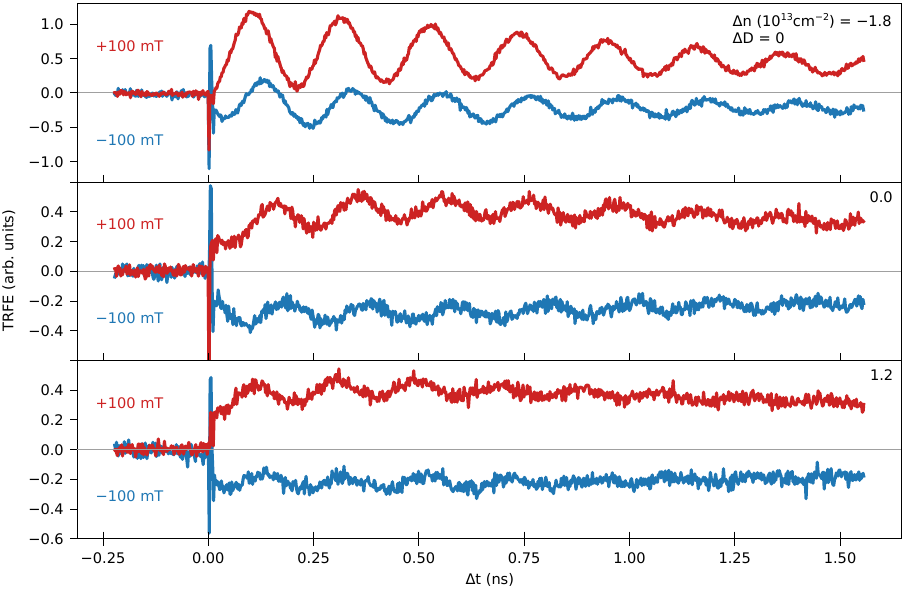}
	\caption{\textbf{TRFE measurements for positive and negative external magnetic fields}.
    Time-resolved Faraday ellipticity measurements at positive and negative external magnetic field ($\mu_0 H_\text{ext} = \pm 100$ mT) and $\Delta D = 0$, for various values of $\Delta n$. The offset at $\Delta t < 0$ is set to zero for clarity. The difference in oscillation amplitude and starting phase for positive and negative $H_\text{ext}$ is clearly visible. Note the different scale for the $\Delta n = -1.8 \times 10^{13}$ cm$^{-2}$ data.
    }
	\label{si_fig:trfe_negative_fields}
\end{figure*}

The measured magnetization precession frequency is found to be independent of the sign of $H_\text{ext}$, (Fig. \ref{si_fig:freq_field-symmetry}).
There are two outliers for $\Delta n$ = 0 cm $^{-2}$ (at -0.75 T and -0.5 T), and one for $\Delta n$ = $1.2\times 10^{-13}$ cm$^{-2}$ (at -0.75 T).
The precession amplitude in the TRFE measurements was too small for these measurements to obtain a good fit.
Fig. \ref{si_fig:freq_field-symmetry}b shows the difference in precession frequency  for positive and negative fields, $f(H_\text{ext}) - f(-H_\text{ext})$.
The absolute difference is smaller than 0.2 GHz, and typically smaller than 0.1 GHz (disregarding the outliers).
Fig. \ref{si_fig:freq_field-symmetry}c shows the relative difference in precession frequency for positive and negative fields, $2[f(H_\text{ext}) - f(-H_\text{ext})]/[f(H_\text{ext}) + f(-H_\text{ext})]$.
Disregarding the outliers, the relative difference is smaller than 2\%, and typically smaller than 1\%.
We point out that especially for $\Delta n$ = $-1.8\times 10^{13}$ cm$^{-2}$, for which the precession amplitude is very large, the frequency difference is very small (0.02 GHz, and 0.2\% for fields above 125 mT).
The TRFE traces at $\mu_0 H_\text{ext} = \pm 100$ mT are shown in Fig. \ref{si_fig:trfe_negative_fields}.
From these measurements it is already clear that reversing $H_\text{ext}$ does not affect the frequency of the oscillations.
These measurements also show again the arguments presented in the main text that the magnetization precession is not purely caused by the $\Delta K$ mechanism: both the amplitude and starting phase of the oscillations are different for positive and negative field.
Furthermore, these measurements also show again the argument presented in the main text and Supplementary Section \ref{si_sec:trfe_ampl_mod} that the gate dependence of the TRFE oscillation amplitude is not (purely) caused by a gate-induced change in the strength of the Faraday effect: the background on top of which the oscillations in the TRFE measurements are, scales differently with $\Delta n$ than the amplitude of the oscillation.

\FloatBarrier
\section{Detailed discussion on the contribution of non-linear magneto-optic effects to the $H_\text{ext}$ asymmetry of the TRFE oscillations.}
Our measurements show that the amplitude of the TRFE oscillations is not symmetric in the external magnetic field.
This could be explained by a change in amplitude of the magnetization precession, as discussed in the main text.
Another possibility is that the amplitude of the magnetization precession is still symmetric in $H_\text{ext}$, but the change in polarization of the probe by magneto-optic effects in the CGT is not.
This could happen if the probe is not only affected by the Faraday effect, which is linear in the magnetization, but also by higher order magneto-optic (MO) effects, such as the Cotton-Mouton effect.
The change in polarization of the probe ($\psi$) due to the MO effects of the sample depends on the sample magnetization as $\psi = a_i M_i + b_{ij} M_i M_j$, where the parameters $a_i$ and $b_{ij}$ are independent of the magnetization, $b_{ij} = b_{ji}$, and $i, j = x,y,z$.
The Faraday effect is described by the first term, and the quadratic MO effect, e.g. Cotton-Mouton effect, by the second term.
In the time-resolved Faraday ellipticity measurements, we observe variations in $\psi$ caused by small variations in the magnetization around the equilibrium magnetization $M_0$.
This is expressed as:
\begin{align}
\Delta \psi = a_i \Delta M_i + 2 b_{ij} M_{0,i} \Delta M_j + b_{ij} \Delta M_i \Delta M_j.
\label{si_eq:nonlinear_mo}
\end{align}

We will now discuss how the quadratic MO effect could affect our TRFE measurements.
The last term in Eq. (\ref{si_eq:nonlinear_mo}) is quadratic in $\Delta \mathbf{M}$, and would result in a double frequency component in the TRFE oscillations.
As this is not observed, this term is assumed to be negligible in our measurements, meaning that $b_{ij}$ and therefore the second order MO effect is negligibly small, or that $\Delta M_i$ is much smaller than $M_{0,i}$

The second term in Eq. (\ref{si_eq:nonlinear_mo}) is linear in $\Delta \mathbf{M}$, and could be of the same order of magnitude as the first term for arbitrarily small values of $\Delta M_i$.
If this is the case, i.e. if $b_{ij} M_{0,i}$ is of the order of $a_i$, this term will result in the oscillation amplitude of $\psi$ being not symmetric in $H_\text{ext}$, even for a thermal induced precession where the oscillation amplitude of $\Delta M$ is symmetric in $H_\text{ext}$.
This is because upon reversing $H_\text{ext}$, $M_{0,i}$ changes sign, changing $\psi$ from $(a_j + b_{ij}M_{0,j}) \Delta M_j$ to $(a_j - b_{ij}M_{0,j}) \Delta M_j$.
Therefore, a quadratic MO effect could result in the observed asymmetry in the amplitude of the TRFE oscillations while the amplitude of the magnetization precession is still symmetric in $H_\text{ext}$.

\begin{figure*}[htbp]
	\centering
	\includegraphics[width=\textwidth]{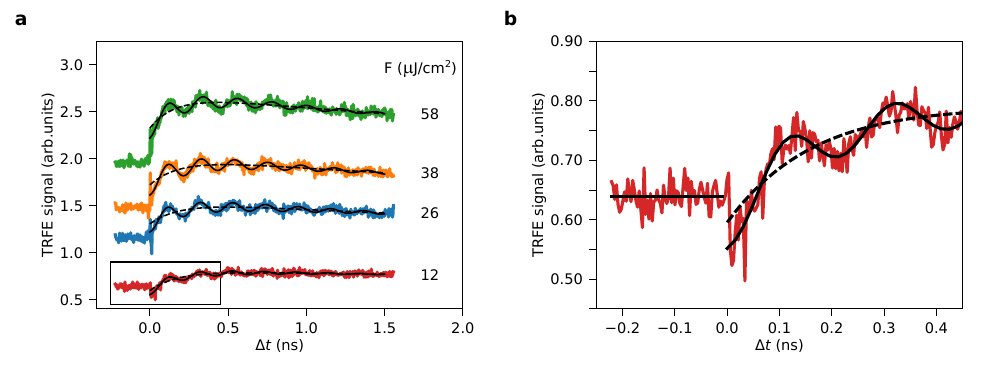}
	\caption{\textbf{Supporting evidence for coherent optical excitation of the magnetization precession.} \textbf{a}, Pump fluence dependence of the TRFE measurements for $V_t = -6V$, $V_b = 6V$ ($D = 1.1$ V/nm, $\Delta n = 0$) and $\mu_0 H_\text{ext} = 100$ mT. The solid line is the best fit of the data, using Eq. (2) of the Methods for $\Delta t > 0$, and a constant for $\Delta t < 0$. The dashed lines indicate the double exponential background of the oscillation
    \textbf{b}, Close-up of the data for a pump fluence of 12 $\mu$J/cm$^{2}$, indicated by the black rectangle in \textbf{a}.  
    }
	\label{si_fig:ICME_argument}
\end{figure*}

However, even though quadratic MO effects could explain that the TRFE oscillation amplitude is asymmetric in $H_\text{ext}$ for a purely thermal excitation of the magnetization precession, they cannot explain all of our observations.
For example, the pump fluence dependence of the TRFE measurements presented in Fig. \ref{si_fig:TRFE_pumppower}e shows a jump in the signal at $\Delta t = 0$ of which the sign depends on the pump fluence.
For convenience, the data is presented again in Fig. \ref{si_fig:ICME_argument}a, with a close-up of the interesting part in Fig. \ref{si_fig:ICME_argument}b.
The data shows that for high pump fluences, the TRFE signal increases sharply right after $\Delta t = 0$, indicating a fast decrease in, or canting of, the magnetization.
For the lowest pump fluence however, the signal sharply \emph{decreases} right after $\Delta t = 0$. The argument for why a purely thermal excitation cannot explain this, as detailed below. 

For a purely thermal demagnetization, the magnetization and the magnetocrystalline anisotropy decrease right after excitation \cite{VanKampen2002}. 
Canting of the magnetization only happens if the magnetocrystalline anisotropy briefly changes on a time scale shorter than the precession period.
This is then accompanied by a a starting phase that is close to $\pm \pi/2$, i.e. if the oscillations start as a sine instead of a cosine (in our experimental geometry) \cite{DallaLonga2008}.
Our measurements however show a clear cosine-like start of the oscillations.
Therefore the magnetization only decreases during the heating by the laser.
Assuming the direction of the magnetization, $M_i / |M|$, has not changed during after the excitation, the TRFE signal can be expressed in terms of the magnitude of the magnetization. Doing this in Eq. (\ref{si_eq:nonlinear_mo}) yields:
\begin{align}
\psi &= a_i M_i + b_{ij} M_i M_j = A |M| + B |M|^2\\
\Delta \psi &= A \Delta |M| + 2B|M_0|\Delta|M|  = \left(A + 2B|M_0|\right) \Delta M.
\end{align}
where $A = a_i M_i/|M|$ and $B = b_{ij} M_i M_j / |M|^2$ are constant.
The sign of $\Delta \psi$ is thus independent of $\Delta M$ and therefore independent of the fluence of the pump.
Hence the change from a sharp increase to a sharp decrease by changing the power of the pump cannot be explained by a pure thermal excitation process.

A coherent excitation on the other hand, caused by e.g. the inverse Cotton-Mouton effect, does have the ability to create an effective magnetic field to cant the magnetization on sub-picosecond timescales \cite{Kalashnikova2008}.
The combination of thermal excitation and coherent laser excitation could explain the observed behavior of the TRFE with pump power.
How this combination can explain the asymmetry in the amplitude and phase with external magnetic field is explained in Supplementary Section \ref{si_sec:coherent_excitations} below.

\FloatBarrier
\section{Model for optical excitation of magnetization dynamics}
\label{si_sec:coherent_excitations}
In this section we develop a model for the excitation of the magnetization dynamics, including coherent optical excitations, and thermal effects.
Using this model, we obtain the amplitude and starting phase of the magnetization precession.

The coherent effects we consider are the inverse Cotton-Mouton effect (ICME) and photo-induced magnetic anisotropy (PIMA).
The effective magnetic field generated by these effects is described by \cite{Yoshimine2014}:

\begin{align}
    H_{\text{eff},i} = G_{ijkl} M_i E_k E_l
    \label{si_eq:imoe_tensor}
\end{align}

\noindent where $G_{ijkl}$ is a rank 4 polar tensor that is symmetric under exchanging the first or last pair of indices, $M_i$ the magnetization, $E_{i}$ the electric field of the light, and $i, j, k, l = x, y, z$.
Note that the two effects are described by two different tensors.
The crystal symmetry of the CGT (space group $R\bar{3}$ \cite{Carteaux1995}) restricts the number of independent elements of these tensors to 12, and forces some to be zero, as indicated in Tab. \ref{si_tab:g-tensor}.
In this table, $G_{ij}$ is an abbreviation of $G_{klmn}$, where i and j take the values 1 to 6, determined by $kl$ and $mn$ respectively. The values for $i$ ($j$) = 1, 2 and 3 correspond to $kl$ ($mn$) = $xx$, $yy$ and $zz$, while the values 4, 5 and 6 correspond to $kl$ ($mn$) consisting of the combination of $y$ and $z$, $x$ and $z$, and $x$ and $y$ respectively.
The 4 boxed elements are responsible for generating an effective magnetic field in the $y$ direction when both the magnetization and the polarization of the pump are in the $xz$ plane.
These elements are needed to explain the the observed asymmetry of the amplitude in the TRFE oscillations for positive and negative values of $H_\text{ext}$.


\begin{table}[htbp]
\begin{tabular}{cl|cccccc}
\hline
\multicolumn{2}{c|}{}                           & \multicolumn{6}{c}{$j$}                                                                                                                                                                                                   \\
\multicolumn{2}{c|}{\multirow{-2}{*}{$G_{ij}$}} & 1                               & 2                                & 3                               & 4                                & 5                                & 6                                            \\ \hline
                                 & 1            & $G_{11}$                        & $G_{12}$                         & $G_{13}$                        & $G_{14}$                         & $G_{15}$                         & $G_{16}$                                     \\
                                 & 2            & {\color[HTML]{656565} $G_{12}$} & {\color[HTML]{656565} $G_{11}$}  & {\color[HTML]{656565} $G_{13}$} & {\color[HTML]{656565} $-G_{14}$} & {\color[HTML]{656565} $-G_{15}$} & {\color[HTML]{656565} $-G_{16}$}             \\
                                 & 3            & $G_{31}$                        & {\color[HTML]{656565} $G_{31}$}  & $G_{33}$                        & {\color[HTML]{CB0000} 0}         & {\color[HTML]{CB0000} 0}         & {\color[HTML]{CB0000} 0}                     \\
                                 & 4            & \framebox{$G_{41}$}                        & {\color[HTML]{656565} $-G_{41}$} & {\color[HTML]{CB0000} 0}        & $G_{44}$                         & \framebox{\color[HTML]{656565} $-G_{54}$} & {\color[HTML]{656565} $-G_{51}$}             \\
                                 & 5            & $G_{51}$                        & {\color[HTML]{656565} $-G_{51}$} & {\color[HTML]{CB0000} 0}        & $G_{45}$                         & {\color[HTML]{656565} $G_{44}$}  & {\color[HTML]{656565} $G_{41}$}              \\
\multirow{-6}{*}{$i$}            & 6            & \framebox{$-G_{16}$}                       & {\color[HTML]{656565} $G_{16}$}  & {\color[HTML]{CB0000} 0}        & $-G_{15}$                        & \framebox{\color[HTML]{656565} $G_{14}$} & {\color[HTML]{656565} $(G_{11} - G_{12})/2$} \\ \hline
\end{tabular}
\caption{\textbf{Symmetry-adapted form of $\mathbf{G_{ijkl}}$.} This table, calculated using Ref. \cite{Gallego2019}, shows the relation between the different elements of $G_{ijkl}$ imposed by the space group $R\bar{3}$ and the invariance under interchanging $i$ and $j$, and $k$ and $l$. using The 12 independent elements are indicated in black, the dependent ones in gray. The elements that are zero are highlighted in red. The boxed elements are responsible for generating an effective magnetic field in the $y$ direction when both the magnetization and the polarization of the pump are in the $xz$ plane. Conversion between $G_{klmn}$ and $G_{ij}$ is explained in the text.}
\label{si_tab:g-tensor}
\end{table}

To determine effect of the ICME and photo-induced magnetic anisotropy on the the precession, we use the following simplified model.
Light travels in the direction $\widehat{\mathbf{k}} = (\sin \theta_k, 0, \cos \theta_k)$, where $\theta_k$ the angle between $\widehat{k}$ and the $z$-axis.
The orientation of the axes is indicated in Fig. 3e of the main text.
The polarization of the pump is linear, and in the $xz$ plane.
Its electric field vector is described by $\mathbf{E} = E_0(\cos \theta_k, 0, \sin \theta_k)$, where $E_0$ is the amplitude of the electric field.
In equilibrium, the magnetization lies in the $xz$ plane too: $\mathbf{M} = M_s (\sin \theta_M, 0, \cos \theta_M)$.
Plugging the above expression in Eq. (\ref{si_eq:imoe_tensor}) reveals that the components of effective magnetic field induced by these coherent optical excitations depends sinusoidally on $\theta_M$ and on $2\theta_k$, where the amplitude and phase are in general different for the different components of $H_{\text{eff},i}$.

When the pump pulse hits the sample, we assume that the magnetization is firstly affected by the ICME.
This effect can be described by a delta pulse in the effective magnetic field, resulting in an instantaneous rotation of the magnetization \cite{Kalashnikova2008, Yoshimine2014, Shen2018}.
Afterwards, the effective magnetic field is changed instantaneously to a new value due to both the PIMA and a thermal induced change in the magnetocrystalline anisotropy.
The former changes the effective magnetic field according to Eq. (\ref{si_eq:imoe_tensor}).
The latter decreases the magnetocrystalline anisotropy of the sample, resulting in a decrease of the effective internal field along the sample normal \cite{VanKampen2002}.
From this point on, the magnetization starts to precess in a circular orbit around the effective magnetic field, as described by the Landau-Lifshitz-Gilbert equation.

The time dependence of $\mathbf{M}$ for negligible damping is conveniently given by Rodrigues' rotation formula:
\begin{align}
    \mathbf{M}(t) = \mathbf{M}_0 \cos\omega t + \left(\widehat{\mathbf{H}}_\text{eff} \times \mathbf{M}_0\right) \sin\omega t + \widehat{\mathbf{H}}_\text{eff} \left(\widehat{\mathbf{H}}_\text{eff} \cdot \mathbf{M}_0\right) \left(1 - \cos\omega t\right)
    \label{eq:m(t)}
\end{align}
which rotates the vector $\mathbf{M}_0$ around a unit vector $\widehat{\mathbf{H}}_\text{eff}$ with angular frequency $\omega$.

The Faraday effect is assumed to be only sensitive to changes in the magnetization along the propagation direction of the laser, $\widehat{\mathbf{k}}$.
The amplitude and phase of the precession as measured by Faraday ellipticity are therefore obtained by projecting the motion of the magnetization on the propagation direction of the laser.
This is achieved by taking the inner product of $\mathbf{M}(t)$ and $\widehat{\mathbf{k}}$.
The resulting expression has a term with time dependence $\cos(\omega t)$, and a term with time dependence $\sin(\omega t)$.
By combining the two into a single phase shifted cosine, the amplitude and starting phase of the TRFE oscillations are obtained.
The amplitude angle ($\theta_A$) of the precession, which is the angle between the magnetization and the effective magnetic field, is calculated too.

For the calculations, we use the experimental parameters $\mu_0\mathbf{H}_\text{int} = 125$ mT and $\theta_H = 50$ degrees.
We set the small thermal induced change of the effective field $\mu_0\Delta H_\text{thermal}$, caused by a reduction of the magnetocrystalline anisotropy and the magnetization, to -1 mT.
A good qualitative agreement with the TRFE oscillation amplitude, and a good quantitative agreement with the starting phase phase, shown in Fig. 3c,d of the main text, are obtained if the ICME and the PIMA depend on the orientation of the magnetization as $\sin\left(\theta_M - 50^\circ \right)$.
The ICME then rotates $\mathbf{M}$ along a fixed direction ($\widehat{\mathbf{H}}_\text{ICME}$), and the PIMA generates an effective field $H_\text{PIMA}$ along a fixed axis, of which the orientations are independent of the value of $H_\text{ext}$.
Furthermore, for the angle over which the ICME rotates $M$ ($\theta_\text{ICME}$), we took $0.4^\circ \sin(\theta_M - 50^\circ)$ along the vector $\widehat{\mathbf{H}}_\text{ICME} = \left(-0.19, 0.96, 0.19 \right)$, and $\mu_0\Delta H_\text{PIMA} = 0.4 \sin(\theta_M - 50^\circ)$ mT along the direction $\left(0.19, 0.96, -0.19 \right)$.

\begin{figure*}[htbp]
	\centering
	\includegraphics[width=\textwidth]{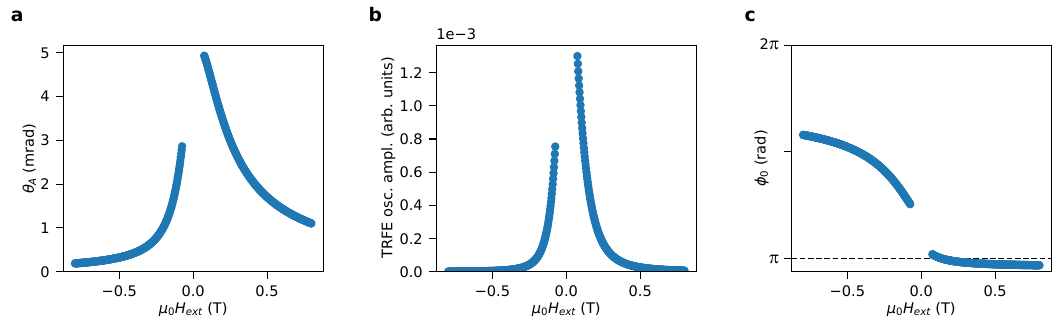}
	\caption{\textbf{Model for the effect of coherent excitations on the magnetization precession.}
    \textbf{a}, The angle between the magnetization and the effective magnetic field versus $H_\text{ext}$.
    \textbf{b}, the amplitude of the percession versus $H_\text{ext}$, as measured by Faraday ellipticity,
    \textbf{c}, The starting phase of the precession versus $H_\text{ext}$.
    }
	\label{si_fig:coherent_excitations}
\end{figure*}

The resulting precession angle, and the TRFE oscillation amplitude and starting phase are shown in Fig. \ref{si_fig:coherent_excitations}.
The values are only plotted for external field magnitudes larger than 75 mT, since at lower values the magnetization is not fully saturated after laser excitation.

The external magnetic field dependence of both $\theta_\text{A}$ and the Faraday ellipticity amplitude are not symmetric in $H_\text{ext}$, and have an inflection point around $\mu_0 H_\text{ext} = \pm 200$ mT.
This agrees with the measured amplitude for $\Delta n$ = $-1.8 \times 10^{13}$ cm$^{-2}$ and $1.2 \times 10^{13}$ cm$^{-2}$.
A possible reason for the discrepancy for $\Delta n = 0$ is given below.
The rate at which the measured amplitude decreases for large fields, and the asymmetry in this rate, is much better captured by $\theta_A$ than by the calculated Faraday ellipticity amplitude.
This could mean that the calculation of the TRFE signal from the magnetization precession is too simplistic in our model model. 

The calculated starting phase ($\phi_0$) also captures the features of the measured starting phase.
For positive fields, $\phi_0$ is mostly constant, at a value slightly below $\pi$, and starts to increase as the field approaches zero.
At negative fields, the phase increases as the magnitude of the field grows, saturating to a value of about $1.7\pi$.
This is in quantitative agreement with our measurements for $\Delta n$ = $-1.8 \times 10^{13}$ cm$^{-2}$ and $1.2 \times 10^{13}$ cm$^{-2}$, and in qualitative agreement for $\Delta n = 0$.

The behavior for $\Delta n = 0$ is very different from to the other two.
The amplitude is mostly symmetric in $H_\text{ext}$ and the relative change with $H_\text{ext}$ for low field is much faster.
Also, the phase appears to be shifted to more positive value of $H_\text{ext}$.
A possible explanation for this is that the thermally induced change in $H_\text{eff}$ does not happen instantaneously, but on a time scale close to the period of the magnetization precession for low fields.
in that case, the amplitude and phase can both be greatly affected if the two time scales are close. 
The strengths of the coherent excitations can also be different than for $\Delta n = 1.2$ or $-1.8 \times 10^{13}$ cm$^{-2}$.

\FloatBarrier
%